**Title:** Demonstration of a laser powder bed fusion combinatorial sample for high-throughput microstructure and indentation characterization

**Authors:** Jordan S. Weaver[1,*], Adam L. Pintar[2], Carlos Beauchamp[3], Howie Joress[3], Kil-Won Moon[3], Thien Q. Phan[1]

[1]Engineering Laboratory, National Institute of Standards and Technology, 100 Bureau Drive, Gaithersburg, MD 20899; jordan.weaver@nist.gov, thien.phan@nist.gov
[2]Information Technology Laboratory, National Institute of Standards and Technology, 100 Bureau Drive, Gaithersburg, MD 20899, adam.pintar@nist.gov
[3]Materials Measurement Laboratory, National Institute of Standards and Technology, 100 Bureau Drive, Gaithersburg, MD 20899; carlos.beauchamp@nist.gov, howie.joress@nist.gov, kil-won.moon@nist.gov
[*]corresponding author

**Graphical Abstract:**

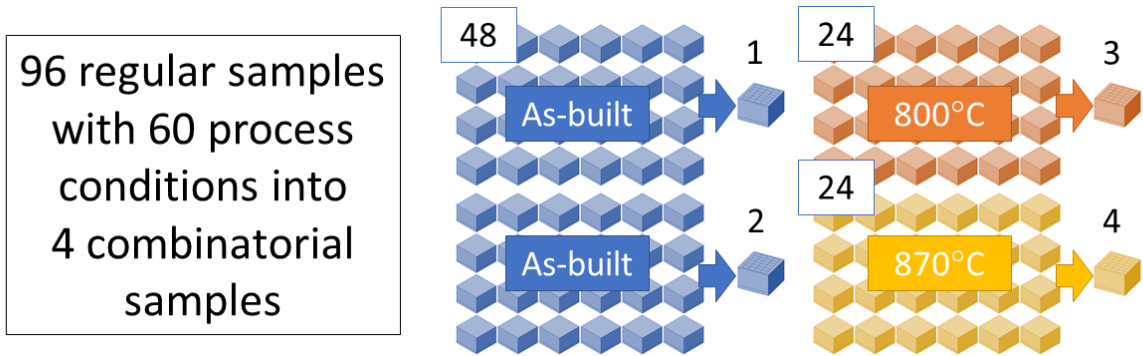

**Abstract:** High-throughput experiments that use combinatorial samples with rapid measurements can be used to provide process-structure-property information at reduced time, cost, and effort. Developing these tools and methods is essential in additive manufacturing where new process-structure-property information is required on a frequent basis as advances are made in feedstock materials, additive machines, and post-processing.  Here we demonstrate the design and use of combinatorial samples produced on a commercial laser powder bed fusion system to study 60 distinct process conditions of nickel superalloy 625: five laser powers and four laser scan speeds in three different conditions. Combinatorial samples were characterized using optical and electron microscopy, x-ray diffraction, and indentation to estimate the porosity, grain size, crystallographic texture, secondary phase precipitation, and hardness. Indentation and porosity results were compared against a regular sample. The smaller-sized regions (3 mm × 4 mm) in the combinatorial sample have a lower hardness compared to a larger regular sample (20 mm × 20 mm) with similar porosity (< 0.03 %). Despite this difference, meaningful trends were identified with the combinatorial sample for grain size, crystallographic texture, and porosity versus laser power and scan speed as well as trends with hardness versus stress-relief condition.

**Highlights:**

- Ninety-six regular samples, sixty unique processes into four combinatorial samples
- Estimated five times reduction in time and cost
- Hardness was most influenced by stress-relieving and sample geometry

## 1. Introduction

Additive manufacturing of metals offers unique advantages for materials and design such as topology optimization and functionally graded materials [1, 2]. Advancing the state of knowledge of process-structure-property relationships in additively manufactured metals is often cited as critically needed for their widespread adoption [3, 4]. Additionally, a more rapid exploration framework is needed due to the time and costs associated with extensive empirical testing [5-8]. High-throughput experiments (HTE) can help to address both needs through automation, combinatorial processing, and rapid measurements [9-14]. Combinatorial processing is the synthesis of a sample library, a single sample that contains sub-samples that vary one or more synthesis parameters (e.g., chemistry and thermo-mechanical processing for alloys), which is conducive to automated and rapid measurements. These methods are often applied in biology and chemistry where thousands of experiments are routine [14]. Within materials science, HTE is applied less frequently and mainly focused on functional materials (e.g., photovoltaic, thermoelectric, energy storage, semiconductor, etc. applications) [15-19]. High-throughput experiments for structural materials (e.g., [9, 10, 20-22]) are even less pragmatic due to challenges in implementing high quality, rapid mechanical tests. The best suited mechanical test for HTE based on speed is instrumented indentation because it measures the location specific mechanical response, is easily automated, and only requires a flat polished surface for testing. However, the property most often used from indentation testing, hardness, is not considered an intrinsic property and can be difficult to correlate with other mechanical properties. Methods to extract more meaningful mechanical properties from instrumented indentation testing is an enduring endeavor (e.g., [23-27]). Aside from indentation, there have been efforts to improve the throughput of traditional mechanical tests, tensile [10] and fatigue [28, 29] testing, as well as develop new tests that capture mechanical performance such as parallel blow forming [30]. Here we note that there are challenges and sacrifices typically made with high-throughput mechanical tests. These include probing small volumes that are less representative of the macroscopic response, activating different deformation mechanisms, and introducing alterations that increase uncertainty compared to standardized test methods. The argument for high-throughput mechanical tests is that for immature processes or new materials, the knowledge gained about the process or material outweighs the deficiencies, and the pursuit will lead to less overall time and effort to achieve the desired performance.

High-throughput experiments applied to AM of metals is promising with several studies demonstrating feasibility and insight into the AM process. For example, Salzbrenner et al. [10] were able to capture the distribution of tensile stress-strain responses for laser powder bed fusion (LPBF) stainless steel (17-4PH) by designing a process to pull arrays of tensile bars in an automated fashion. Gong et al. [31] used instrumented spherical indentation stress-strain curves to characterize chemically graded Ti-Ni alloy produced via directed energy deposition (DED) for generating process-structure-property relationships. Knoll et al. [32] applied rapid alloy prototyping methods (RAP) to tool steels produced via DED, measuring hardness and tensile properties to tune strength and ductility. Lastly, Shao et al. [28] outlined a framework for using ultrasonic fatigue testing to accelerate the design of AM Ni superalloys. The current study differentiates from those listed above by developing a combinatorial sample that contains various laser parameters rather than focusing on feedstock or alloy design. Samples that contain various laser parameters are not new, but arguably underutilized for high-throughput experiments. For example, Ahmed et al. [33] resourcefully printed a combinatorial sample with 5 laser scan speeds to study part porosity for different power re-use conditions. The part was not leveraged for further process-structure-property information and only varied a single laser parameter.

Design of experiments (DoE) has been successfully applied in LPBF to screen for the most influential variables, perform process optimization, and build models using both simulations and experiments. For example, Delgado et al. [34] used a 2-level full factorial design to experimentally study the effects of scan speed, layer thickness, and building direction on iron based alloys, Ma et al. [35] used thermal finite element simulations to determine critical variables for nickel superalloy 625, Kamath et al. [36] optimized laser process parameters for fully dense 316L stainless steel at high laser powers, and Calignano et al. [37] used a full factorial design to develop a regression model for fully dense thin wall structures as a function of the laser process parameters. Applying DoE to LPBF is an open area of research as evidenced by efforts such as Gheysen et al. [38] who compared full factorial and central composite designs while exploring good settings of laser power and scan speed to minimize porosity, and Mahmoodkhani et al. [39] who showed the benefits of a modified Plackett-Burman design compared to factorial designs. Similar to DoE, the objective of high-throughput experiments may be screening for the most important variables, building regression models, discovering new materials, and/or process optimization. To draw the distinction, high-throughput experiments with combinatorial processing is a complimentary method to DoE; it can help to complete a designed experiment with less effort or run a larger designed experiment that is not practical with the conventional approach of many individual samples.

The focus of the present study is to design and demonstrate a combinatorial, high-throughput approach for laser powder bed fusion that includes laser process parameters and post-process parameters. The hypothesis is that a combinatorial, high-throughput approach will produce meaningful process-structure-property information while reducing the time, cost, and effort compared to a more traditional approach. Process-structure-property information for a more traditional approach will be pulled from literature and supplemented with a limited number of measurements on regular samples. Section 2 describes the combinatorial sample, characterization, and data analysis. Section 3 presents the results, which are divided into microstructure and indentation. Comparisons to trends in literature and a regular sample are also provided for each measurement in this section. Section 4 discusses strategies for improving the method based on the results as well as the time and cost savings for the combinatorial approach. This is followed by the conclusions in Section 5.

## 2. Materials and methods

### 2.1 Experiment design and combinatorial sample

Nickel superalloy 625 was chosen because there is extensive literature on how the LPBF laser parameters and post-processing influence the part performance. In particular, laser power, laser scan speed, and stress-relief temperature are commonly studied. The three main variable settings (laser power, scan speed, and condition) in the experiment design are listed in Table 1. Condition is used to describe as-built and post-processing heat-treatments. In Table 1, laser power and scan speed are given as a percentage of the default setting. In principle, the changes in laser power and scan speed mimic thermal history variations that may occur during AM resulting in under heating and overheating. Laser power ranges from 68 % to 130 %. A value of 130 % is at the maximum laser power for the AM machine (370 W). Laser scan speed ranges from 83 % to 146 %. The choice of two levels at higher speeds compared to one level at a lower speed was made since increasing build time with faster speeds is of general interest. Note that laser power and scan speed settings of 80 % and 125 %, respectively, were specifically chosen because of their use in on-going round robin studies organized by NIST. The 20 combinations of laser power and scan speed were printed in a single sample, referred to as a combinatorial sample, containing 24 columns with 4 mm × 3 mm cross-sections as shown in Fig. 1. Four extra columns were used to print repeats at the default settings. Columns 1, 10, 16, and 19 were purposely chosen for repeats at the default settings because they span diagonally across the sample, which allows for a first order check of spatial dependence and neighborhood affects (i.e., possible thermal affects from surrounding columns). The remaining column numbers were randomly assigned the 20 combinations of laser power and speed, which are listed in Table 2. Four combinatorial samples were printed with the same configuration of column numbers and settings. Two combinatorial samples were characterized in the as-built condition, and two separate samples were prepared for heat treating by sealing in evacuated ($< 2.7 \times 10^{-6}$ Pa) quartz ampules. One sample was heat treated at 800 °C for 1 hour (SR 800 °C) and the other at 870 °C for 1 hour (SR 870 °C). Both samples were rapidly cooled (ampule sealed water quenching). The latter heat treatment, SR 870 °C, is recommended by the manufacturer; however, it produces approximately 2 vol.% of δ-phase [40-42]. The first heat treatment, SR 800°C, was shown to reduce the δ-phase to < 0.5 vol.% [40-42]. The four combinatorial samples create a total of 96 samples with 60 unique process conditions.

**Table 1. Three variables and their levels in the design of experiments. Default laser power and speed are 295 W and 960 mm s$^{-1}$, respectively. All other process parameters were fixed at that default settings and are listed in the supplementary material.**

| Laser power (% default) | 68, 80, 100, 111, 130 |
|---|---|
| Laser scan speed (% default) | 83, 100, 125, 146 |
| Condition | As-built, Stress-relieved 1 hour at 800°C (SR 800°C), Stress-relieved for 1 hour at 870°C (SR 870°) |

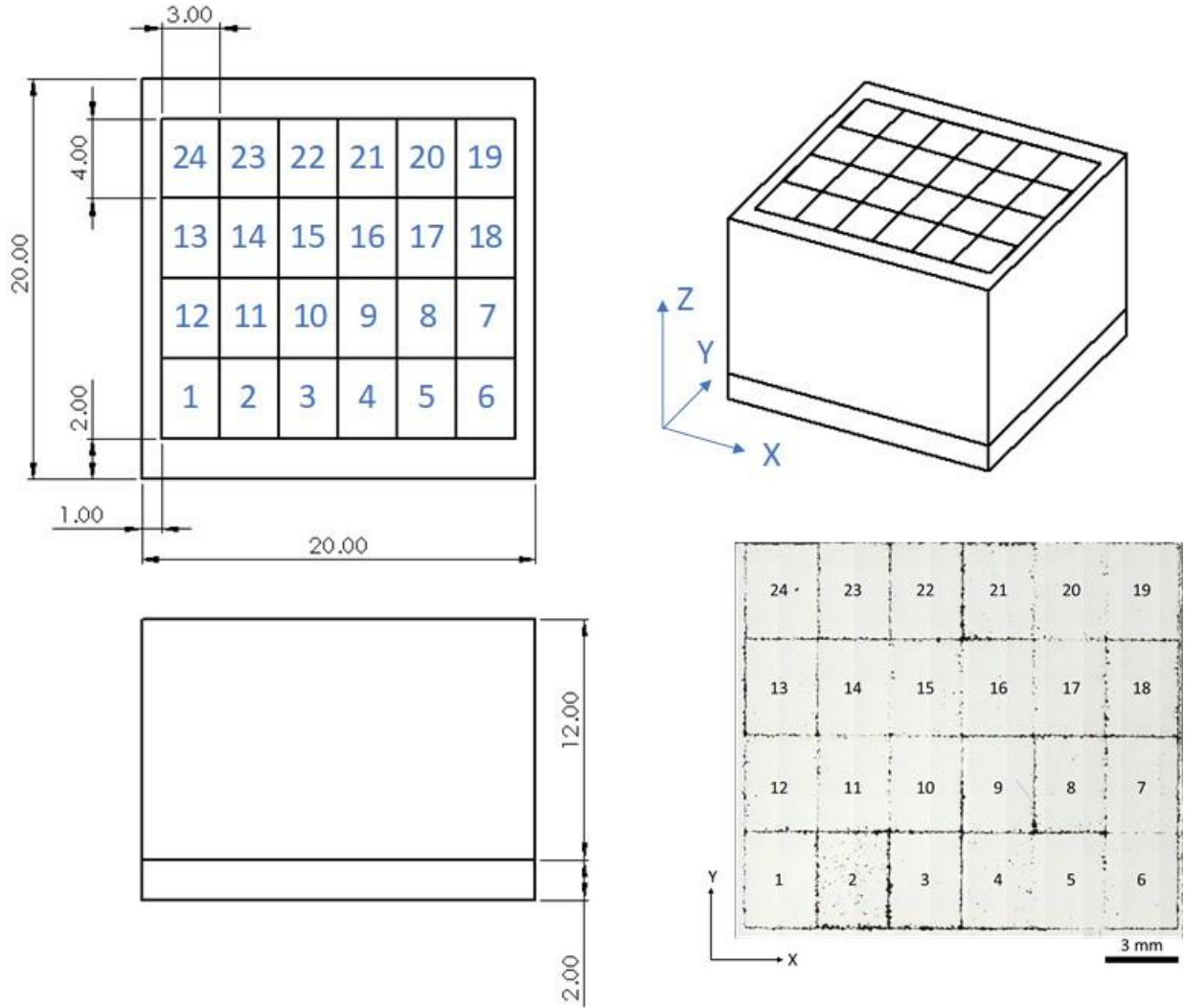

**Fig. 1 Drawing and dimensions of combinatorial sample. Units are in mm. Cross-sectional optical micrograph of the printed part for the Z plane. The laser process parameters for each column number are given in Table 2. Positive Z is the build direction and negative X is the recoating direction. The surrounding shell and base are printed with default settings. The print order starts with the shell and fills in the columns from 1 to 24.**

**Table 2 Combinatorial prescribed laser power and scan speed. There are five replications of the default setting. The maximum and minimum power to speed ratios are noted.**

| Column Number | Power (W) | Scan Speed (mm s$^{-1}$) | Note | Column Number | Power (W) | Scan Speed (mm s$^{-1}$) | Note |
|---|---|---|---|---|---|---|---|
| 1 | 285 | 960 | Default | 13 | 228 | 700 | |
| 2 | 195 | 1400 | Min. Ratio | 14 | 316 | 960 | |
| 3 | 195 | 960 | | 15 | 370 | 1200 | |
| 4 | 370 | 700 | Max. Ratio | 16 | 285 | 960 | Repeat |
| 5 | 370 | 960 | | 17 | 285 | 1400 | |
| 6 | 285 | 700 | | 18 | 228 | 1200 | |
| 7 | 316 | 700 | | 19 | 285 | 960 | Repeat |
| 8 | 228 | 1400 | | 20 | 370 | 1400 | |
| 9 | 285 | 1200 | | 21 | 195 | 1200 | |
| 10 | 285 | 960 | Repeat | 22 | 195 | 700 | |
| 11 | 316 | 1400 | | 23 | 228 | 960 | |
| 12 | 316 | 1200 | | 24 | 285 | 960 | Repeat |

There were additional considerations in the design of experiment. Rather than use a traditional factorial or fractional factorial design, the three factors under study each have a different number of levels: five levels of laser power, four levels of laser scan speed, and three levels of condition or stress relief. More than two levels for laser power and scan speed are necessary because the relationship between the response (microstructure or hardness) and those variables may be curved, or possibly even non-monotonic. The design of experiment is a split-plot design instead of fully randomized because the combinatorial sample is a contiguous unit that is built and stress-relieved as such. Split-plot designs are commonly used to reduce the burden of randomization [43], and they are often necessary in additive manufacturing for variables that are fixed for each build or run (e.g., powder layer thickness, powder lot, chamber gas, etc.). Lastly, the sample was dimensionally constrained to fit inside a typical metallographic mount, 31.75 mm in diameter, so it could be easily inserted into existing machines for preparation and characterization. Non-symmetric features (rectangular column cross-section and variable shell thickness) were purposely used to make it easier to distinguish in-plane directions.

**2.2 AM process and sample characterization**

A commercial laser powder bed fusion (LPBF) machine (EOS[1] M290) was employed with settings previously listed in Table 1 and Table 2. In addition to the four combinatorial samples, regular samples with the same dimensions of the combinatorial unit (20 mm × 20 mm × 12 mm) were built at the default settings for comparison. The feedstock material was re-used nickel superalloy 625 powder (EOS IN625) sieved with a 63 µm mesh size. The chemical composition of parts was measured and conforms to the ASTM standard specification for additive manufactured UNS N06625 with powder bed fusion (see supplementary material for the measured chemical composition) [44]. Samples were removed from the build platform using wire-electric discharge machining (EDM) approximately 1 mm from the build platform and cross-sectioned horizontally with a precision saw approximately 5 mm from the top of the sample. All characterization was done in the plane approximately 5 mm from the top of the sample. Note that this sample design focuses on the Z-plane, and modifications and other design considerations would be required to study a transverse cross-section.

Cross-sectioned samples were metallographically prepared with a final vibratory polish using 0.02 µm colloidal silica for various microstructure measurements. Where noted, samples were etched with Aquia Regia [45] by immersion for approximately 30 seconds. The microstructure was characterized with optical and electron microscopy. Optical microscopy was performed on a Zeiss LSM 800 microscope. Optical micrographs were taken with a field of view 2.634 mm × 1.756 mm and pixel resolution of 1.86 µm per pixel. Image processing was performed using FIJI [46] for porosity and pore shape analysis. First, images were cropped in the x-direction by 10% on each side to avoid material near the edges of each column. Next the images were segmented using the auto (default) threshold, which is a global histogram algorithm using iterative intermeans followed by a pore size analysis using the Analyze Particles plugin to determine the percent area fraction and circularity of the pores. The minimum pore size was limited to three pixels. Electron microscopy was performed on a JOEL JSM7100F field-emission scanning electron microscope (SEM). Additionally, electron backscatter diffraction (EBSD) was performed with an Oxford NordlysMax[2] detector using a beam voltage of 20 keV, step size of 2.7 µm, 4 × 4 camera binning, and tilt angle of 62°. The lower than 70° tilt angle was necessary to be able to move each column in the sample under the SEM pole piece without having to rotate the sample. The EBSD data were analyzed using MTEX, a free MATLAB toolbox for analyzing and modeling crystallographic textures by means of EBSD or pole figure data [47]. Grains were defined by a minimum misorientation angle of 3° and minimum size of 9 pixels (each pixel corresponds to a measurement with a dimension equal to the step size of 2.7 µm), which results in a minimum grain size equivalent radius of 4.6 µm. The misorientation angle of 3° captures subgrains in the grain size distribution that were missed with the default value of 5°. Small clusters of non-indexed data less than 10 pixels in size were assigned to surrounding grains. This is less than 5 % of the data, and in most cases less than or equal to 2 % of the data. The orientation distribution function (ODF), necessary for texture analysis, was calculated using the mean orientation per grain weighted by the grain area [48, 49]. The texture index, the $L^2$-norm of the ODF, and the entropy of the ODF were used to track changes in the overall texture, which provide indices for the texture strength and deviation from a uniform texture, respectively [50]. Lastly, X-ray diffraction (XRD) was performed on a Bruker D8 diffractometer using Cu-K$\alpha$ radiation over a 2$\theta$ scan range of 40º to 55º, a step size of 0.01º, and acquisition rate of 400 s per degree of $2\theta$ on the SR 870 °C sample. This was done to confirm the presence of δ-phase and determine the relative differences of δ-phase. The diffraction peak intensities were calculated as the integral of intensity versus the scattering vector ($q$) after subtracting off a linear background. The cross section of the X-ray beam was approximately 0.5 mm in diameter to fit inside each 3 mm × 4 mm region, even at a low incident angle. Characterization of dislocation cells, $\gamma'/\gamma''$ precipitates, Laves phases, and chemical segregation are often critical for understanding structure-mechanical property relationships in nickel superalloys. However, this was not included in the current study due to a lack of rapid tools and/or to limit the scope for a proof-of-concept study.

---

[1] Certain commercial equipment, instruments, or materials are identified in this paper in order to specify the experimental procedure adequately. Such identification is not intended to imply recommendation or endorsement by the National Institute of Standards and Technology, nor is it intended to imply that the materials or equipment identified are necessarily the best available for the purpose.

    Nano and microindentation were used to measure the modulus and hardness. Nanohardness measurements were made using a Keysight G200 nanoindenter with a diamond Berkovich tip. The continuous stiffness measurement (CSM) was used, which imposes a small oscillatory signal (2 nm displacement amplitude at 45 Hz) to generate many small elastic unloads necessary to determine the modulus and hardness. A constant strain rate method was employed at a value of 0.05 $s^{-1}$. The speed at which the tip approaches the sample was increased to as high as 50 nm $s^{-1}$ compared to the default value of 10 nm $s^{-1}$ in order to reduce the time per indent. The modulus and hardness were determined following the Oliver-Pharr analysis [51] at a displacement range of 475 nm to 500 nm. A sample Poisson's ratio of 0.31 was used to determine the sample modulus [52]. The tip area function was determined on a fused quartz reference sample. Because of the large number of nanoindents, indents on a fused quartz reference sample were performed periodically to confirm the modulus and hardness was not changing significantly. If a significant deviation was detected, the tip area function was re-calculated based on the reference sample data. The average measurements on quartz and area

function coefficients are provided in the supplementary material. A total of 3,325 nanoindents on AM combinatorial samples were performed with approximately 20 tests per column on stress-relieved samples and approximately 90 tests per column on as-built samples (see supplementary material). Additionally, Vickers hardness measurements were made following ASTM E92 [53] with a test force of 4.903 N (0.5 kgf). Approximately nine tests per column on each sample were made resulting in 644 indents on combinatorial samples. In addition to the combinatorial samples, nano and micro indentation measurements were performed on a regular sample made with the default settings for direct comparison to the five columns in the combinatorial sample that were also built with the default settings.

A summary of the samples and characterization is provided in Table 3, which lists the different measurement types along with a description of the statistical analysis that was applied to generate confidence intervals. To account for the split-plot nature of the experiment as well as repeated measurements on each column, the hardness measurements (nano and micro) were modeled with a linear mixed effects model leveraging the `lmer` function from the lme4 package [54] of R [55] for estimation. Confidence intervals (95% level) for mean hardness at each condition, power, and speed combination were constructed using a parametric bootstrap algorithm. The porosity and pore shape measurements were also modeled using a linear mixed effects model, and confidence intervals (95% level) were created for the mean porosity and pore shape of each laser power and speed combination leveraging a parametric bootstrap algorithm. The grain size analysis included between 550 and 1100 grains per column on one as-built combinatorial sample. Each sample was used to estimate the quantiles of the grain size distribution for its column. The quantiles and their corresponding 95% confidence intervals were estimated nonparametrically using the default method of the `eqnpar` function in the EnvStats package [56] of R [55]. One measurement of texture index and entropy per column were made on one as-built combinatorial sample. The mean texture index and entropy, for each combination of laser power and speed was estimated using an analysis of variance (ANOVA) model. Confidence intervals (95% level) are the standard *t*-intervals for ANOVA models. The `lm` function of R [55] was used for the calculations. The data and calculations for $\delta$-phase are the same as for texture index and entropy, except that $\delta$-phase was measured on the SR 870 °C combinatorial sample instead of the as-built combinatorial sample.

**Table 3. Matrix of samples and measurements with methods for error estimates. Filled circles indicate all columns were measured in the combinatorial sample. A partially filled circle indicates a partial number of columns were measured in the combinatorial sample. The sample size range per column is listed in the last column. Note that the sample size for the default condition is generated from five columns (i.e., approximately five times more).**

| Measurement Type | As-built #1 | As-built #2 | SR 800 °C | SR 870 °C | Confidence Intervals | Compared to regular sample | Sample Size Range |
|---|---|---|---|---|---|---|---|
| Nanohardness | ● | ● | ● | ● | Mixed Linear Model with Parametric Bootstrap for Intervals | Yes | 18 – 93 indents |
| Vickers hardness | | ● | ● | ● | Mixed Linear Model with Parametric Bootstrap for Intervals | Yes | 7 – 9 indents |
| Porosity and pore shape (Optical) | ● | | ● | | Mixed Linear Model with Parametric Bootstrap for Intervals | Yes | 14 – 283 pores |
| Grain size and texture (EBSD) | ● | | | ◐ | Texture: ANOVA Model with Student-*t* Intervals Grain Size: Non-parametric Estimates and Intervals | No | 569 – 1052 grains |
| δ-phase (XRD) | | | | ● | ANOVA Model with Student-*t* Intervals | No | 1 scan |

## 3. Results

### 3.1 Microstructure

The results in this section and those that follow are organized in terms of their settings (i.e., laser power, laser scan speed, and their ratio) for observing trends. Some plots are based on column numbers when observing variation within the combinatorial sample and comparing the combinatorial and regular samples for the default laser process conditions. Figure 2 shows optical micrographs of the as-built sample organized into a 5 × 4 matrix of laser power along the vertical axis and laser scan speed along the horizontal next to a 5 × 1 array of all the positions built at the default settings. In other words, there are twenty distinct combinations of laser power and scan speed and four repeats at the default settings with one image duplicated (with an asterisk) for comparison. The field of view width (2.634 mm) is only slightly smaller than the fiducial grid (every 3 mm) so that porosity at the edge of the images should be ignored. The laser power to scan speed ratio can be used as a coarse estimate for how the thermal history changes [57]. Increasing the linear heat input (power to speed ratio) increases the peak temperature and decreases the cooling rate. The actual thermal history depends on more factors, and no attempt was made to estimate or model the thermal history for the combinatorial process conditions. At the extremes of low power and high speed (bottom right of array in Fig. 2) there are significant lack of fusion defects. In the other extreme of the array at high power and low speed (top left in Fig. 2) there is spherical porosity that is likely due to keyholing or a more unstable metallurgical process. Additional micrographs from a second sample are provided in the supplementary material, which shows similar trends, along with the segmented images and a table of the results from both samples. Figure 3 shows the 20 porosity and circularity mean estimates and corresponding confidence intervals (one for each laser scan speed and power combination) versus the ratio of laser power to scan speed, which support the aforementioned observations. That is, at a low ratio in the lack-of-fusion range, mean porosity increases above 0.1 % with pores that are less circular than those in the default settings, and at a higher ratio, there is an increase in mean porosity above 0.1 % with pores that remain circular in shape. Note that these values should be considered coarse estimates since best practice dictates using several micrographs from multiple planes at higher magnifications [58-61].

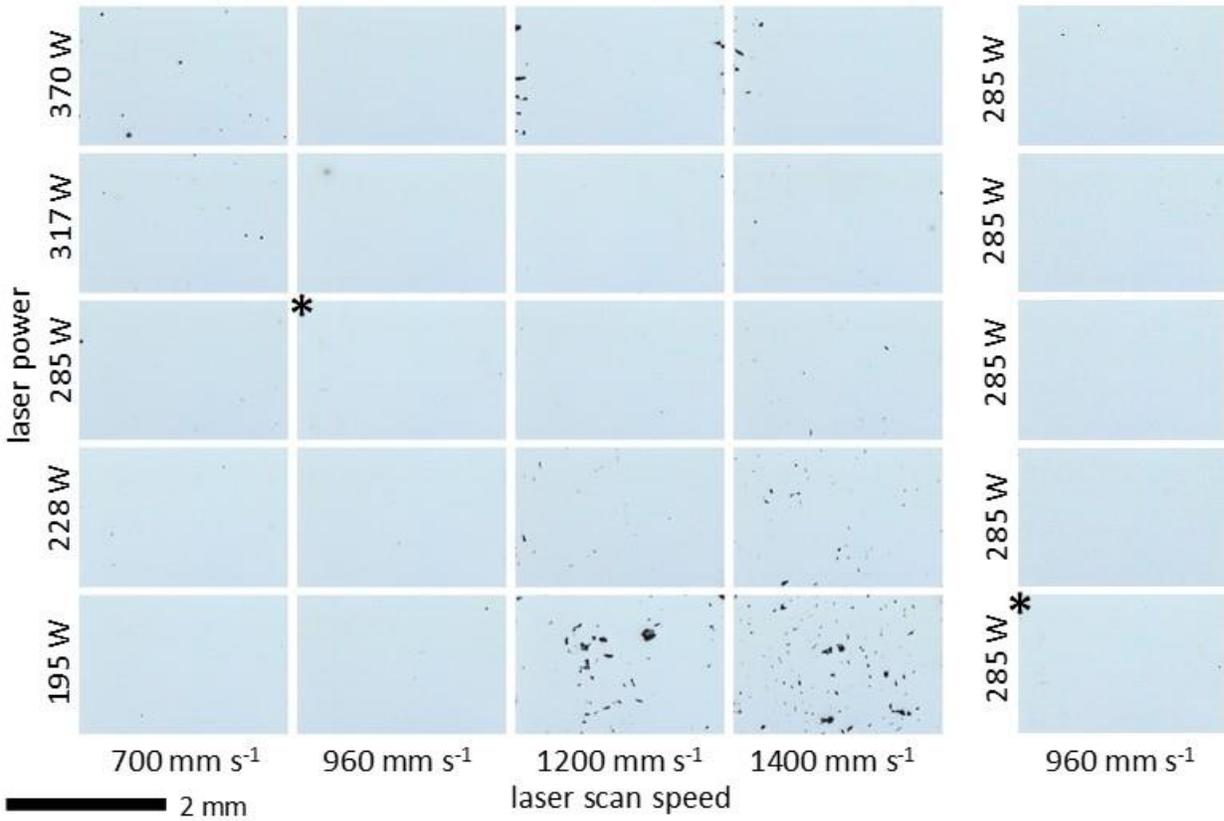

**Fig. 2 Optical images of combinatorial sample sorted by laser settings and repeats. The grid in this format is only 4 × 5, which combined with the top 4 rectangles in the last column produce the 4 × 6 grid shown in Fig. 2.**

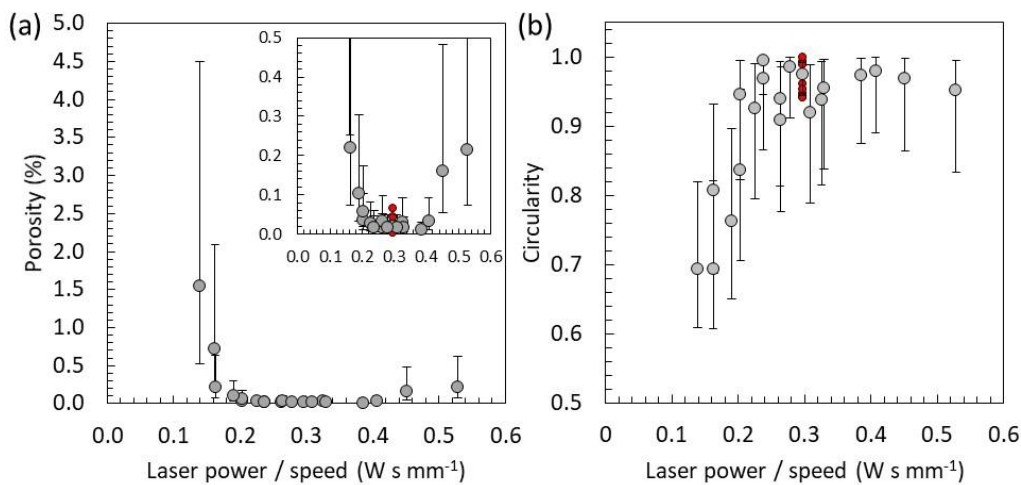

**Fig. 3 (a) Mean percent porosity versus the ratio of laser power to scan speed with an inset rescaled to better show low porosity data points (b) Mean pore circularity versus the ratio of laser power to scan speed. Porosity (area fraction) was determined by one micrograph per column on two different combinatorial samples. See Table 3 for a description of the interval calculations. Smaller, red data points are the five columns at the default settings.**

Figure 4 shows porosity measurements for the five repeated columns (1, 10, 16, 19, 24) in the combinatorial sample and the regular (R) sample, which are all fabricated with the default laser power and scan speed. Since there are duplicates of each column (as built and 870 °C), and the default laser power and scan speed are repeated for five columns, if the post processing condition is ignored (a reasonable assumption for porosity), the role of column can be assessed in the observed variability in the measurements. We find that it is not a significant contributor to the observed variability and so conclude that there is not a significant trend of porosity versus column number (or column position) after accounting for laser power and speed. That conclusion matches what is seen in Figure 4. Figure 4 also shows porosity measurements from a single regular sample are in reasonable agreement with the combinatorial sample for the default laser parameters. This is also important as it demonstrates similarity between the small columns in the combinatorial sample and regular sample.

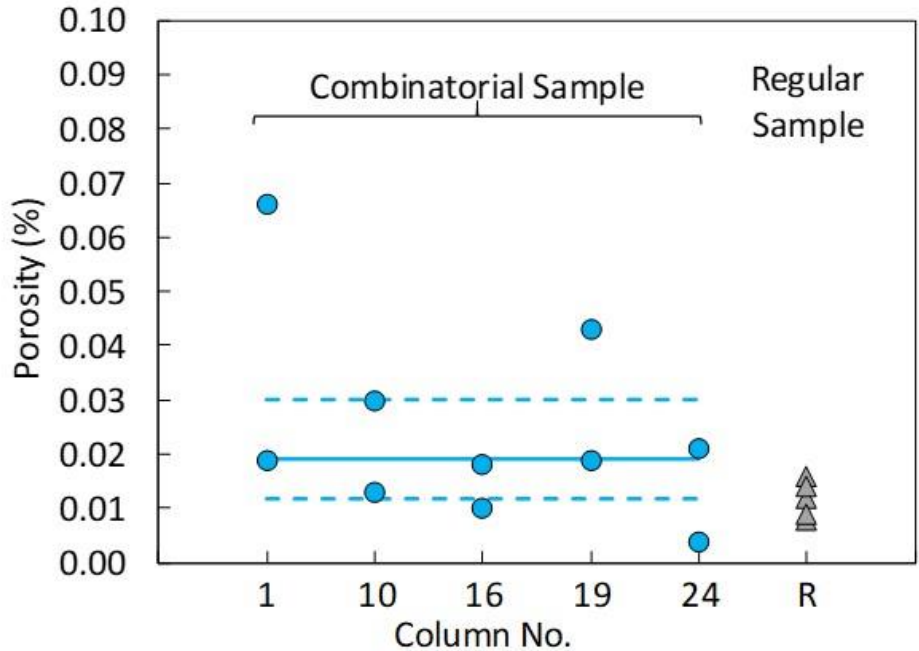

**Fig. 4 Porosity measurements versus column number (refer to Fig. 1) from combinatorial samples for the default laser power and scan speed compared against porosity measurements from a regular sample (R) for the default laser power and scan speed. The solid and dotted lines are the estimate and confidence interval determined for the default condition in the combinatorial sample.**

In the same format as Fig. 2, the crystallographic microstructure is shown in Fig. 5 by electron backscatter diffraction (EBSD) measurements. It is apparent that the microstructure becomes finer with increasing laser scan speed. Note that black in the raw band contrast images, Fig. 5a, is caused by grain boundaries and pores. The white space in the processed inverse pole figure (IPF) maps, Fig. 5b, are regions with no crystallographic information caused by pores, poor indexing (grain boundaries and high local misorientation) as well as grains < 9 points that were not included in the analysis. Additionally, it is important to note that these micrographs are from the build plane and not representative of the material as a whole. The LPBF process often produces elongated grains along the build direction (not shown here) [4]. Measurements from multiple planes that account for the non-equiaxed grain shape are required to describe the 3D grain size and morphology (e.g., [62, 63]). Nevertheless, there is highly useful information about the process from a single cross-section of the combinatorial sample. Figure 6 shows the as-built grain size measurements determined from Fig. 5 versus the laser power to scan speed ratio. The grain size distribution was estimated by the empirical cumulative distribution function to produce the equivalent radius at the $5^{th}$, $50^{th}$, and $95^{th}$ percentiles. Figure 6 conveys that the median ($50^{th}$ percentile) grain size decreases slightly with decreasing laser power/speed ratio, and the $95^{th}$ percentile (i.e., the larger grains) decreases significantly with decreasing laser power/speed ratio. This is explained in part by slower cooling rates at higher ratios that allows for more grain growth time [4, 57]. Esmaeilizadeh et al. [62] report a decrease in grain size with increasing laser scan speed for LPBF nickel superalloy Hastelloy X, which is consistent with this work. A select number of EBSD data was collected on the SR 870 °C sample, see supplementary material, which indicate that the grain size distribution did not change significantly from the as-built to stress-relief conditions. Thus, the grain size measurements from Fig. 6 are representative of all three conditions in this study. Grain growth at 870 °C may be limited by precipitates at grain boundaries. For example, Zener pinning by Nb rich precipitates has been observed in nanocrystalline nickel superalloy 625 [64].

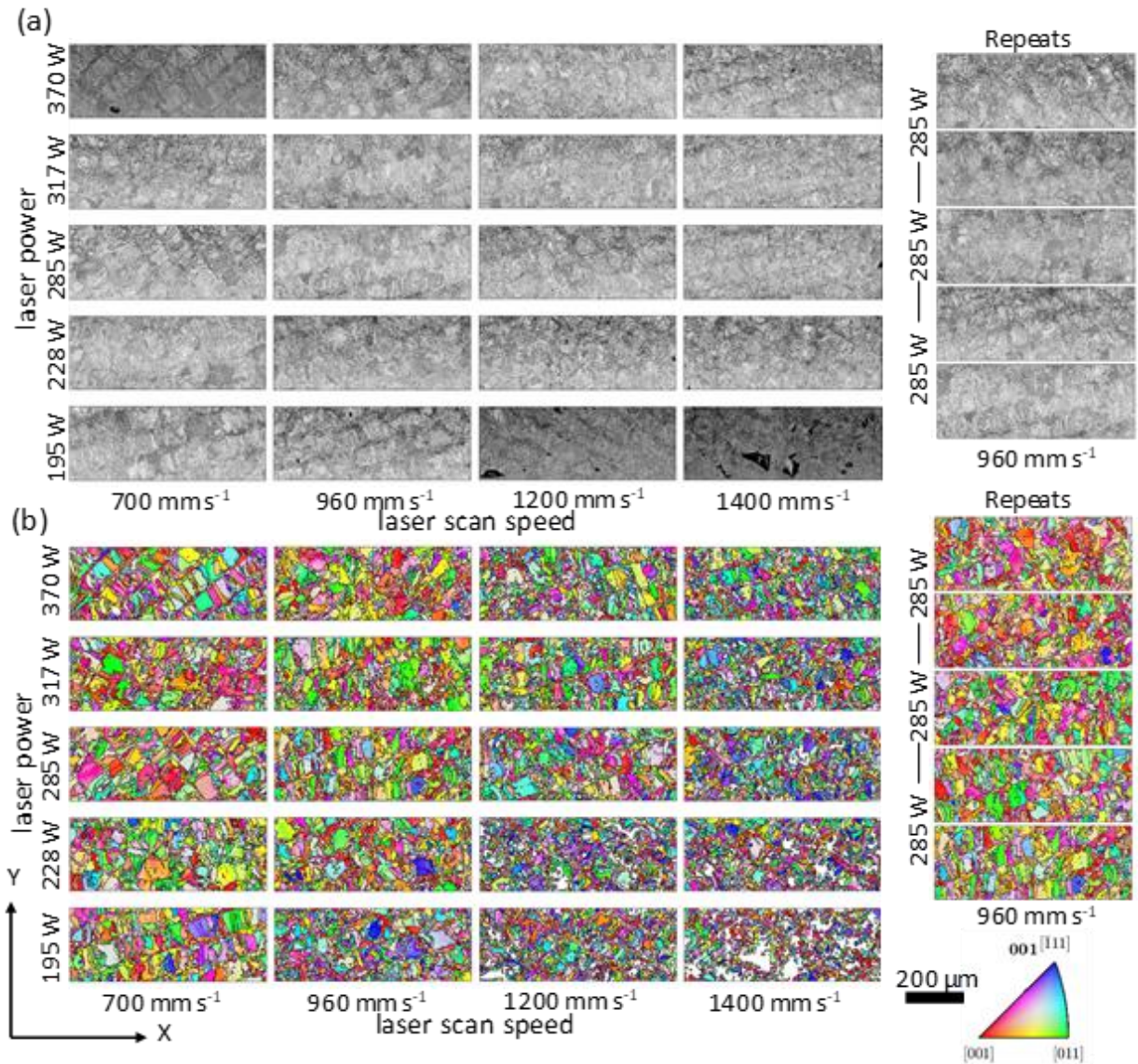

**Fig. 5 EBSD (a) band contrast and (b) inverse pole figure maps of as-built combinatorial sample. The field of view for each micrograph is approximately 0.948 mm × 0.354 mm. The scale bar applies for all images.**

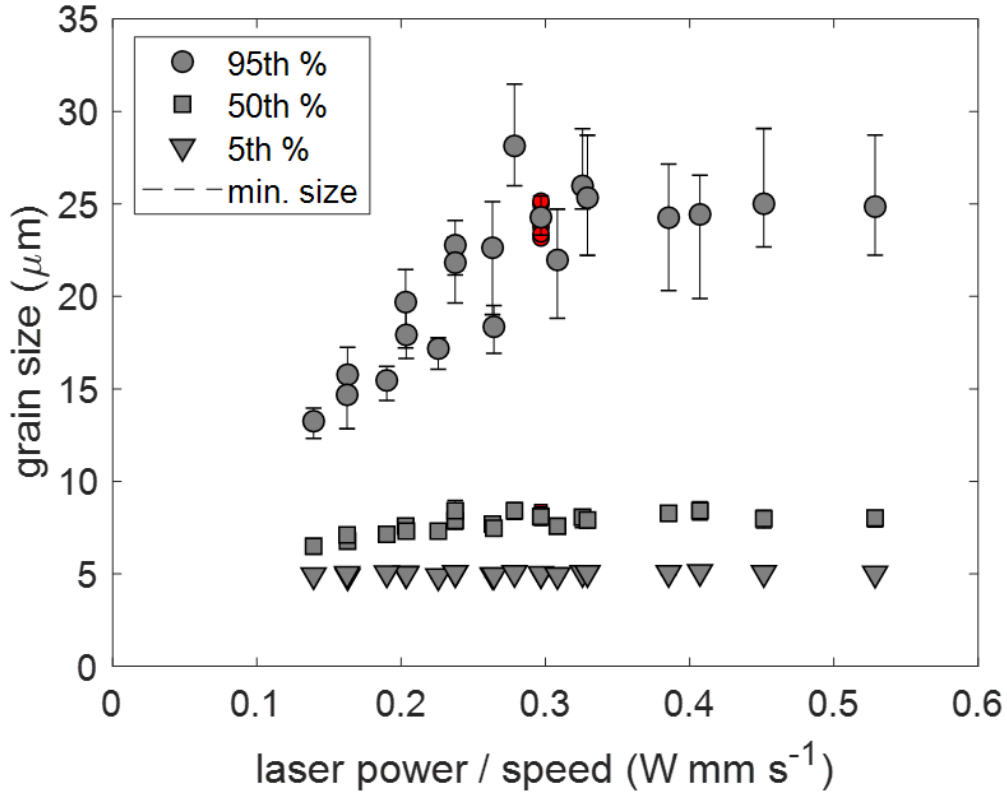

**Fig. 6 Grain size distribution percentiles (%) for as-built combinatorial sample. The number of grains in the measurements per column ranged from 560 to 1052 with a mean of 758 grains. See Table 3 for a description of the error bar calculations. Note that the 5th percentile is unreliable because it is very close to the minimum detectable grain size (min. size) of 4.7 μm. Smaller, red data points are the five columns at the default settings.**

Crystallographic texture plots obtained from EBSD are shown as IPF contour maps in Fig. 7. The IPF texture map illustrates which crystal planes are parallel to the build direction. Regions of high intensity (e.g., greater than 2 multiples of a uniform distribution) indicate a preferred crystal orientation. A value of 1 multiple of a uniform distribution (mud.) indicates a uniform distribution of crystal orientations or the absence of texture. The most prevalent texture is a near {100} texture, which is commonly reported for LPBF nickel superalloy 625 [41, 65]. It appears that the texture goes from moderate to no distinct texture (fully uniform or random) with increasing laser scan speed. This is supported by the texture index and entropy of the ODF [50] plots in Fig. 8, which show values approaching uniform crystal distributions with decreasing laser power/speed ratios. Choo et al. [66] observed a similar change from a strong (100) texture to random in LPBF 316L stainless steel for a fixed laser scan speed and decrease in laser power. Here we note that there is variation in the texture measurements in Fig. 7 and Fig. 8 for the default settings. Larger area scans containing more grains are required to produce better statistics for texture analysis to confirm the trends. The number of grains for accurate ODFs and texture measurements ranges from thousands to 10's of thousands depending on the ODF and accuracy required [67, 68].

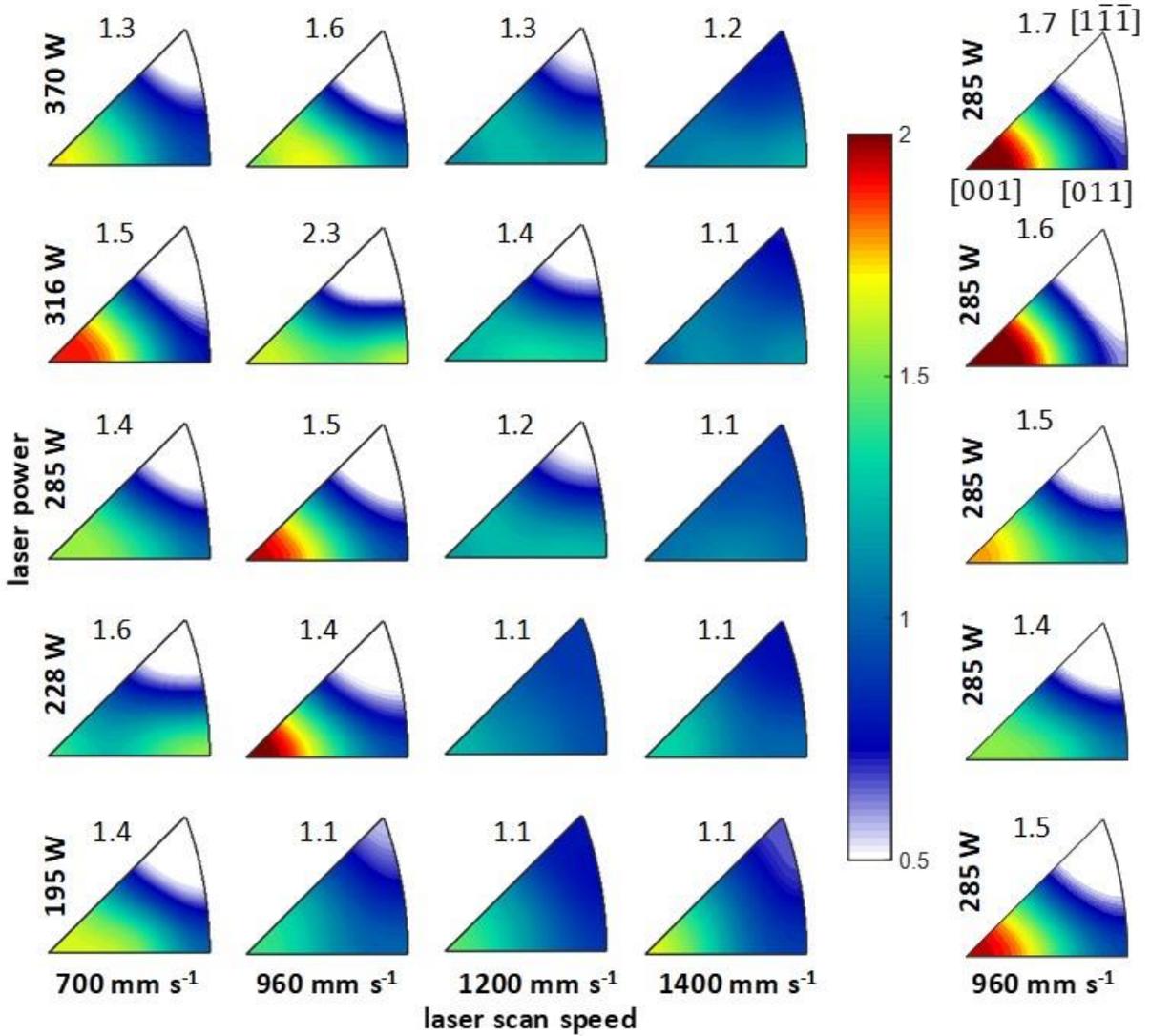

Fig. 7 Crystallographic texture plots for as-built combinatorial sample. Units for the color bar are multiples of a uniform distribution (mud). A value of 1 mud indicates no preferred texture. The texture index, $L^2$-norm of the orientation distribution function, is listed for each dataset. Similarly, a texture index of 1 corresponds to a uniform distribution with no preferred texture. The texture index for the repeats at the default settings has a mean (standard deviation) of 1.5 (0.1).

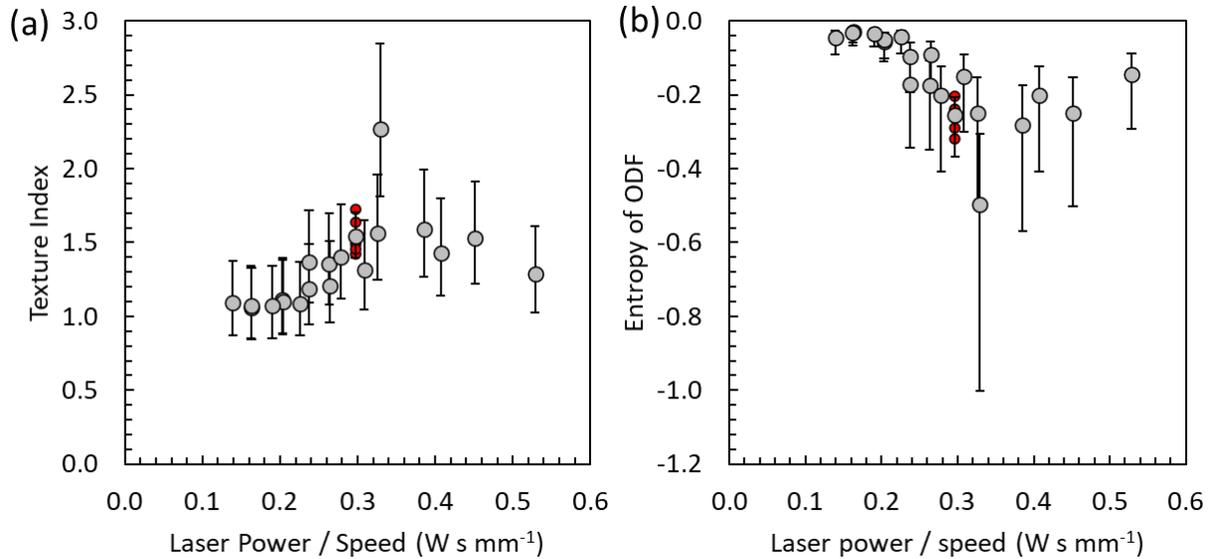

**Fig. 8 (a) Texture index versus the ratio of laser power over scan speed. A texture index of 1 indicates a uniform distribution or no texture. (b) Entropy of the orientation distribution function (ODF). An entropy of 0 indicates a uniform distribution or no texture. See Table 3 for a description of the interval calculations. Smaller, red data points are the five columns at the default settings.**

Figure 9 shows scanning electron microscopy (SEM) micrographs of etched samples for all three conditions at select laser parameters (default, lowest laser power/speed ratio, and highest laser power/speed ratio). Several studies have detailed the precipitation of δ-phase in stress-relieved LPBF nickel superalloy 625 (e.g., [40-42]). Under 2D projection, the plate-like δ-phase precipitates have a needle-shaped appearance, as seen in Fig. 9 for SR 870 °C. The root cause for this is the chemical segregation of Nb and Mo to dendrite walls, which decreases the energy required to form δ-phase. δ-phase readily forms at 870 °C for 1 hour (SR 870 °C) but remains very low when annealed at 800 °C for 1 hour (SR 800 °C). XRD was used to confirm the presence of δ-phase for the SR 870 °C sample (Fig. 10a). The δ-phase peak intensity (formed by reflections of the {211} and {012} crystal planes) can be used as an analog for the relative changes in volume fraction of the δ-phase as laser power and scan speed change. This assumes a fixed texture for the δ-phase, which is a decent assumption because of the small δ-phase grain size. Figure 10b shows that there is no trend of the δ-phase peak intensity versus the laser power/speed ratio. This suggests that the different laser parameters in this study do not lead to differences in the formation of δ-phase after stress-relief.

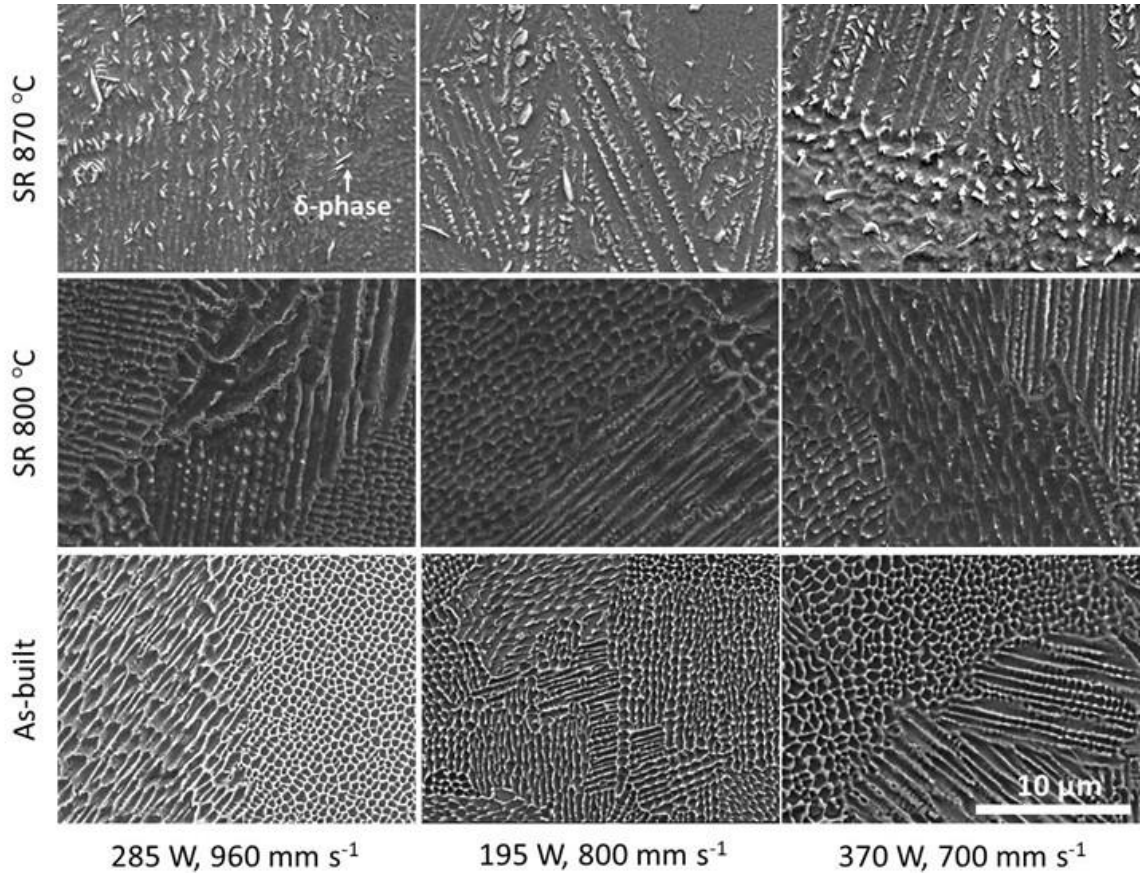

**Fig. 9 SEM micrographs of etched samples for three conditions and three laser power and laser scan speed combinations default, lowest power over speed ratio, and highest power over speed ratio. The rows are the three conditions, and the columns are the three power-speed combinations.**

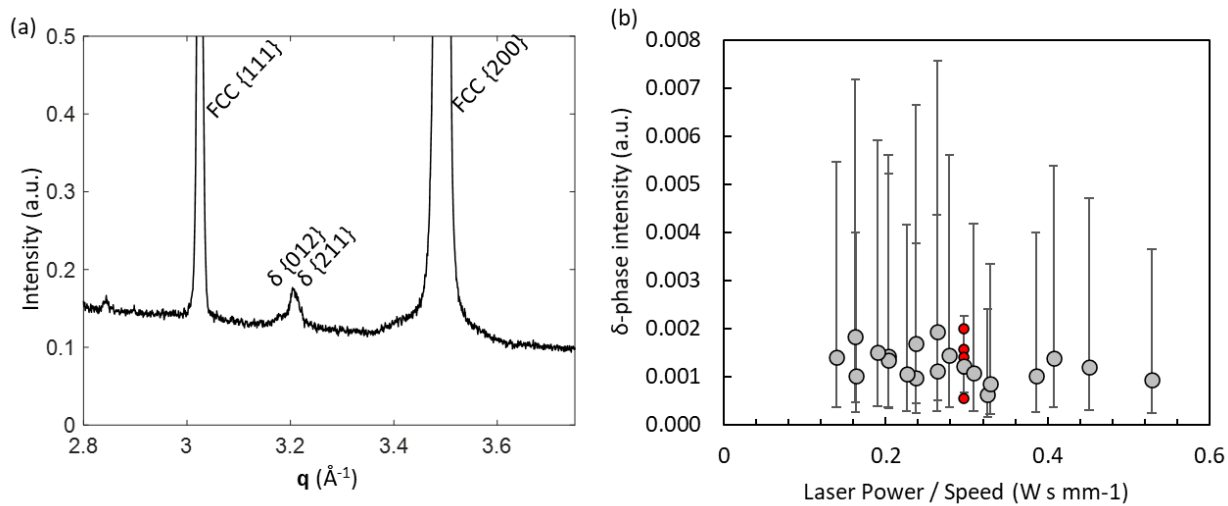

**Fig. 10 (a) Representative XRD profile on the SR 870 °C sample for a 2$\theta$ scan of 40° to 55°. $q$ is the scattering vector (b) δ-phase peak intensity versus laser power / scan speed ratio. See Table 3 for a description of interval calculations. Smaller, red data points are the five columns at the default settings.**

## 3.2 Indentation

Nanoindentation modulus and hardness and microindentation hardness measurements are presented in Fig. 11 for the repeated columns in the combinatorial sample in the as-built condition. Again, these repeated columns were built with the default laser parameters in specific positions for a first order estimate of neighbor affects (refer to Fig. 1 and Table 2). A comparison of the mean of the five columns to a regular sample is also provided in Fig. 11. First, there is no significant trend in the indentation measurements based on column number or position within the combinatorial sample. In one case, the hardness is different in one column compared to the majority of the other columns: Column 24 Vickers hardness (Fig. 11e) has a lower hardness than Columns 10, 16, and 19. The exact influence of neighbor effects requires more tailored experiments; however, the use of repeated columns within the combinatorial samples serves as a coarse way to determine the extent of neighbor effects. Except for the above-mentioned case, these neighbor effects appear to be minimal. Second, the comparison between the average of the repeated columns to a regular sample shows that there is a difference in hardness. The modulus is similar for the combinatorial and regular sample means (lower bound / upper bound): 205.7 GPa (201.1 / 209.8) and 211.8 GPa (210.9 / 212.8), respectively. However, the nano and microindentation hardness in the combinatorial sample is lower, 4.19 GPa (4.18 / 4.21) and 291.4 HV 0.5 (288.4 / 294.5), compared to the regular sample, 4.82 GPa (4.80 / 4.84) and 309.5 HV 0.5 (304.8/ 314.3), respectively. This indicates that the geometry of the columns (4 mm $\times$ 3 mm area) in the combinatorial sample compared to the regular sample (20 mm $\times$ 20 mm area) has a significant effect on hardness. Here we note that the geometry of the regular samples is an arbitrary choice. A complex part will have thin and thick cross-sections that are not necessarily represented by a single "regular" sample. For the purpose of trends within the combinatorial sample, the difference between a regular sample and combinatorial samples does not matter. However, the influence of sample geometry is an important aspect in the design of the combinatorial sample, which will be discussed more later.

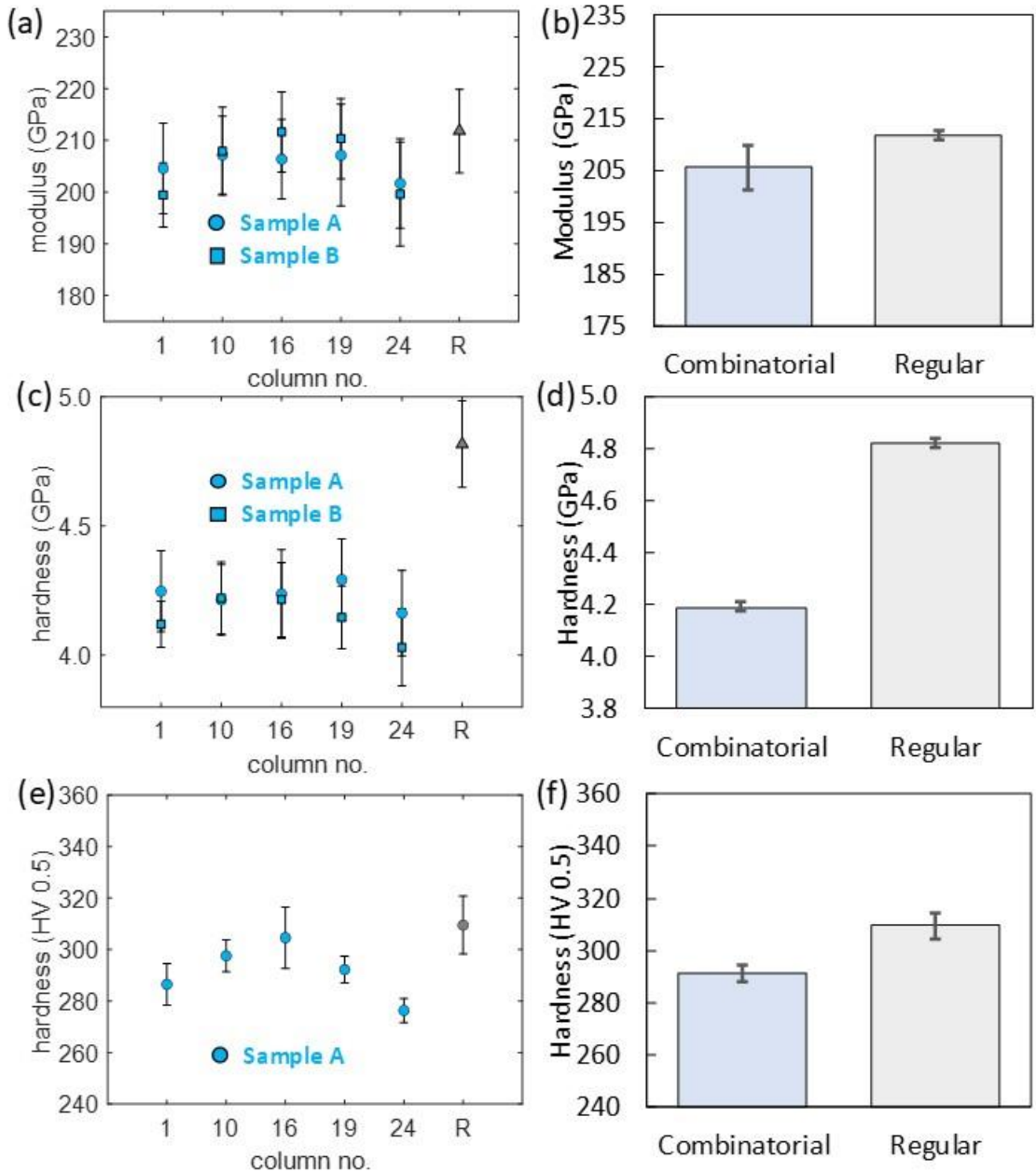

Fig. 11 Comparison of indentation measurements for default laser parameters on combinatorial columns and a regular sample in the as-built condition. (a, c, and e) have individual column averages ± one standard deviation compared to the regular (R) sample average ± one standard deviation. Two samples (Sample A and B) were included for nanoindentation measurements. (b, d, and f) compare the expected values and their 95% confidence intervals. The combinatorial sample confidence intervals were determined according to Table 3. The regular sample confidence intervals were determined using Student-t intervals for $n = 280$ and $n = 24$ for nanoindentation and microindentation, respectively.

Nanoindentation modulus and hardness and microindentation hardness measurements are plotted against the laser power to scan speed ratio in Fig. 12. In addition, the hardness measurements are plotted versus laser scan speed for the default laser power (285 W) in Fig. 13. Extra plots of the data based on laser power and speed are included in the supplementary material. The modulus, nanoindentation hardness, and microindentation hardness do not show significant trends with the ratio of laser power to scan speed, which is in contrast to changes in porosity, grain size, and crystallographic texture. It's noted that the microindentation hardness shows a slight decreasing trend for decreasing laser power to scan speed ratio below a value of 0.2. The nanoindentation size (a cross-sectional area equivalent radius of approximately 1.38 µm at maximum load) and microindentation size (cross-sectional area equivalent radius of approximately 22 µm) are smaller than or equal to the grain size (see Fig. 6), which means they will not capture a change in hardness due to a change in grain size. The indents are also not sufficiently large enough to reliably sample the effect of porosity. In most cases, indents that hit a porosity defect fail to run or cannot be analyzed, which is particularly true for nanoindentation. The Vickers indents provide a better sampling volume for porosity; however, they are arguably too small because they are less than the size of an individual laser scan track (see Supplementary material for a representative micrograph). The slight drop in Vickers hardness for a low laser power to scan speed ratio is likely caused by indents that land near porosity and not directly sample porosity defects. Arrays of indents smaller than the grain size still sample the effect of crystallographic texture; however, the texture changes were mild and the hardness dependence on crystal orientation is moderate for FCC crystals [69-71]. Thus, it is reasonable that there is no significant hardness trend versus the laser power to scan speed ratio. The main factor affecting the hardness was the condition (as-built, SR 800°C, and SR 870°C), which can be explained by considering precipitation hardening.

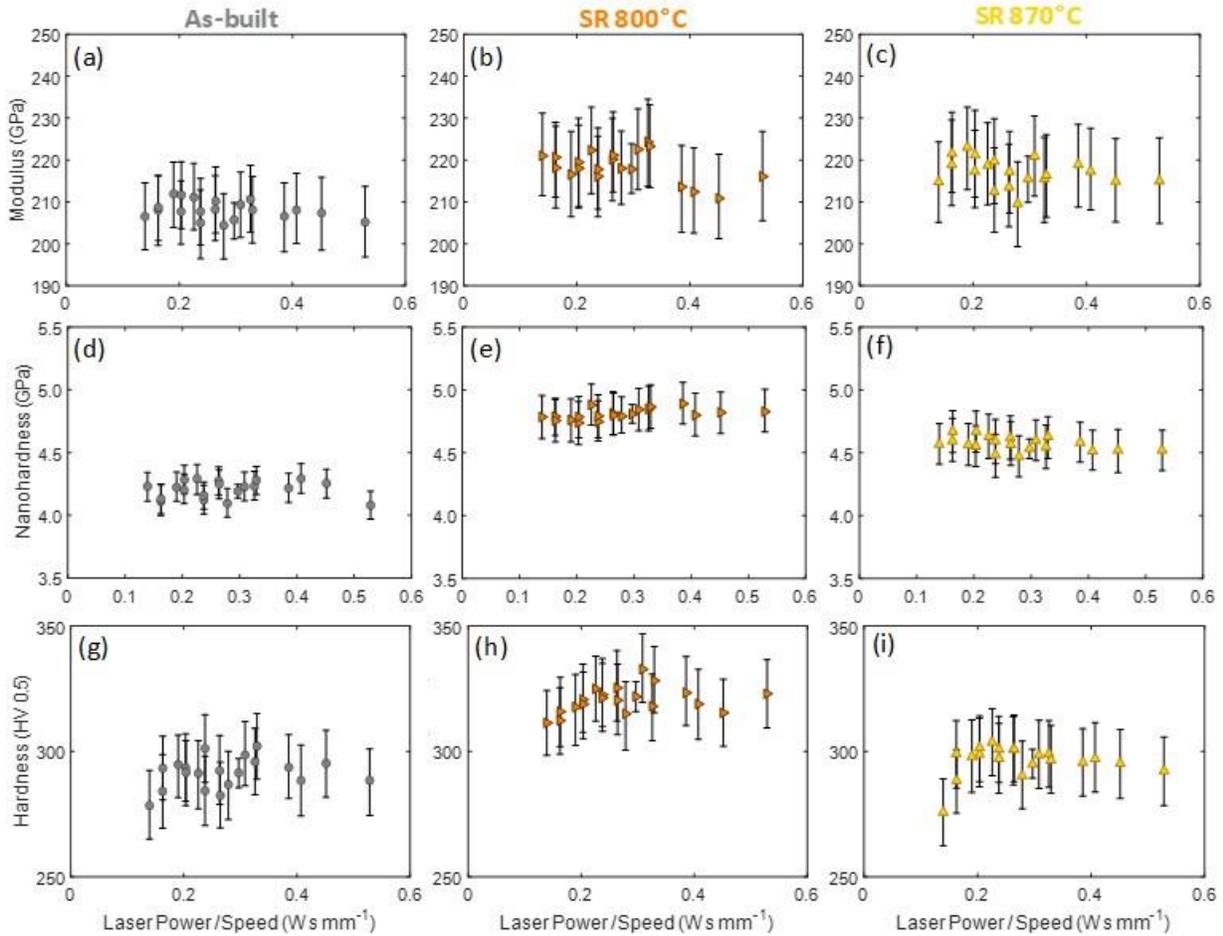

**Fig. 12 Nanoindentation modulus and hardness and microindentation hardness for rows 1, 2, and 3 respectively. The horizontal axes for every plot are the ratio of laser power to scan speed. Columns are separated by condition: as-built, SR 800 °C, SR 870 °C. See Table 3 for a description of the interval calculations. A complete table of the mean, upper, and lower bounds is provided in the supplementary material.**

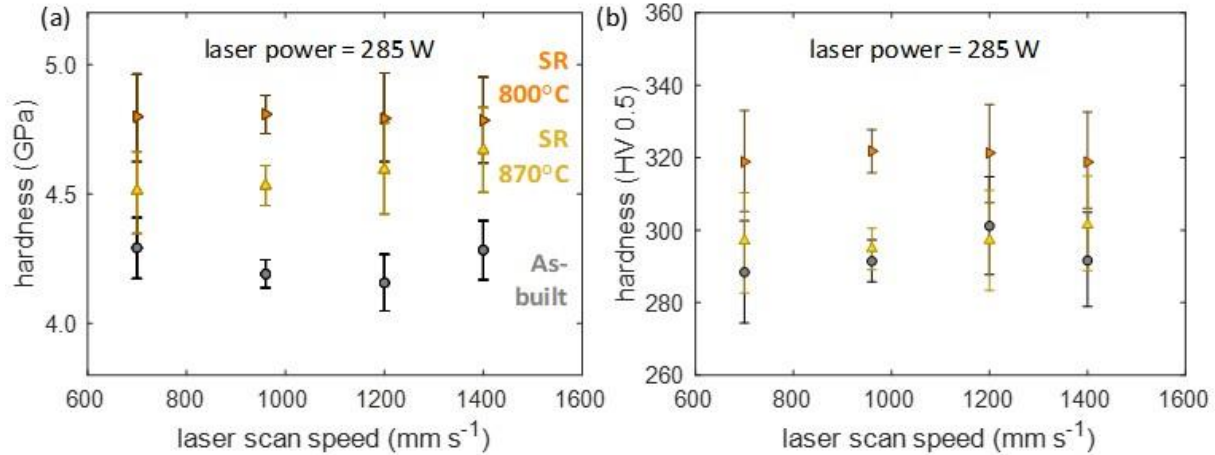

**Fig. 13 (a) nanoindentation hardness versus laser scan speed and (b) microindentation versus laser scan speed for a fixed laser power of 285 W. See Table 3 for a description of the interval calculations.**

The hardness increases from as-built to SR 800°C, and then it decreases from SR 800°C to SR 870°C. For example, the nanoindentation hardness is approximately 4.2 GPa, 4.8 GPa, and 4.5 GPa for the as-built, SR 800 °C, and SR 870 °C, respectively. In this case, the microstructure features that can significantly influence hardness are dislocations and precipitates. These were not characterized in this study due to the lack of rapid measurement tools (e.g., TEM). An explanation is given based on findings reported in literature. Stress-relief heat treatments typically reduce the dislocation density or have no effect on the dislocation cell structures (e.g., LPBF IN718 [72]) that formed during AM, which would result in a drop in hardness or no change in hardness. Hardness actually increased after stress-relief in this study. During stress-relief, precipitation hardening (resulting in an increase in hardness) can also occur, and the size, morphology, and type of precipitates greatly depend on the stress-relief time and temperature and degree of chemical micro segregation at dendrite boundaries. Carbides may be present in the as-built condition [73, 74] and can increase with stress-relief heat treatments along with the precipitation of other carbides (MC, $M_2C$, $M_6C$ and $M_{23}C_6$) [41, 42, 74-76]. Furthermore, Laves phases have been observed in L-PBF nickel superalloy 625 [73]. All these phases can increase the yield strength and hardness, although $\gamma'$ and $\gamma''$ are the most potent due to their strained coherent interface with the matrix [74, 77-79]. Marchese et al. [73] studied a solutionizing heat treatment followed by aging at various times and temperatures for LPBF IN625. They observed an increase in hardness from as-built after 2 hours of aging at 700 °C and 800 °C followed by a drop in hardness for aging at 900 °C. This agrees with the hardness trends in this study: an increase in hardness after stress-relief with a higher hardness for 1 hour at 800°C compared to 1 hour at 870°C. In this work, only δ phase was identified using XRD. The volume fractions of other precipitates were too small to determine with lab-based XRD. Additionally, some diffraction peaks of $\gamma'' - BCT$ are shared with $\gamma - FCC$, making it difficult to identify with XRD. Again, we rely on observations in literature to explain the trend in hardness, which we hypothesize is caused by the process of $\gamma''$ precipitation followed by $\gamma''$ turning into δ and/or the coarsening of fine precipitates. Observations of $\gamma''$ turning into δ with continued aging have been reported in conventional IN625 [80-82]. Additionally, Suave et al. [79] observed high hardness values associated with precipitation of $\gamma''$ that dropped when $\gamma''$ started to turn into δ-phase in conventional IN625. In LPBF IN625, Lass et al. [83] identified the presence of $\gamma''$ along with δ after stress relief at 870 °C for 1 hour. Since there is less δ-phase for a stress relief at 800 °C compared to 870 °C [76, 83], it is expected that there will be more $\gamma''$ in the SR 800 °C compared to SR 870 °C. Thus, the precipitation of $\gamma''$ during stress-relief with a higher fraction of $\gamma''$ (lower fraction of δ) for 1 hour at 800 °C compared to 1 hour at 870 °C would result in the highest hardness for SR 800°C. This hypothesis is consistent with hardness trends with stress-relief time and temperature in this study and literature, but it requires careful TEM characterization to validate.

## 4. Discussion

Some brief explanations of structure-property relationships were given with the results. The discussion here focuses on the design and use of the combinatorial sample for high-throughput characterization. Strategies for improving the method regarding the following are detailed: column size, neighbor affects, indentation methods, and cost/time savings.

### 4.1 Design of combinatorial sample

The column cross-sectional dimensions, 4 mm × 3 mm, were smaller than the default stripe width of 10 mm (maximum laser track length before turning around) and smaller than a more typical 20 mm × 20 mm area for metallographic samples. This resulted in a higher hardness in the regular sample for the default laser power and scan speed. One possible explanation for this is that the difference in the time between neighboring laser tracks is enough to create different thermal histories (i.e., cooling rates). In the small area in each region of the combinatorial sample, the laser never travels more than 5 mm in distance (the diagonal across the 3 mm × 4 mm area) before doubling back. Often it travels much less before coming back to re-melt some of the previous track. In the regular sample, the maximum laser travel is not limited by the sample area; it travels up to 10 mm in distance set by the stripe width. The length of each laser track will vary just as in the smaller area, but overall, the significant difference between the small and large sample would lead to faster cooling rates and higher temperature gradients in the larger regular sample. Such a difference could lead to differences in microstructure and hardness. Various sample sizes should be considered for combinatorial samples and regular samples. A possible combinatorial design would include two levels for the cross-sectional area with a large and small area. This along with the stripe width, which both control the maximum laser scan length, will be considered in a future study.

Thermal affects from neighboring columns within the combinatorial sample were anticipated. As a first step, repeated columns (columns built with the same laser process parameters) were placed in specific locations within the combinatorial sample so that if the affect was significant, it would show up as significant differences between these columns. The microstructure and hardness results indicate that the affect is minimal. This can partly be explained by the lack of fusion porosity between columns demarking the column boundaries, see Fig. 1, which provide some thermal isolation. For example, 90 % porous aluminum foams have an effective thermal conductivity one order of magnitude less than the solid material [84]. An improved design to reduce heat transfer between columns would be to create thicker porous dividers or leave a gap of unmelted powder since the powder bed typically has an order of magnitude lower thermal conductivity than the solid material [85]. The only drawback to this improvement is that it can lead to a larger sample or fewer columns for a fixed sample size. Optimizing the design to prevent neighbor affects while maximizing usable sample space could be achieved with thermal modeling.

**4.2 Indentation methods**

The indentation methods (Berkovich nanoindentation and Vickers microindentation) were chosen due to their prevalence and versatility to make many indents inside small regions. This was at the expense of surveying larger volumes of material. This resulted in hardness measurements that were not sufficiently large enough to establish microstructure-property relationships for changes in grain size and porosity. Indenting to higher loads or depths resulting in larger contact area will increase the interaction volume under the indenter. Using a spherical-conical tip can further increase the interaction volume such as Rockwell or Brinell hardness measurements. Hardness measurements provide a low effort, rapid measurement to indicate differences in resistance to plastic deformation. Indentation measurements can also be used to extract static mechanical properties such as the uniaxial yield strength, ultimate tensile strength, or the engineering stress-strain curve [23-25, 27, 86-89]. These more advanced indentation methods rely on using fully instrumented indentation, finite element simulations of the indentation test, and matching experiments and simulated load-displacement, effective stress-strain, residual indent profiles, and additional parameters. These methods require significant cost up front to establish a sound basis for converting indentation measurement into uniaxial measurements. The assemblage of combinatorial samples and advanced indentation protocols for mesoscale process-structure-property relationships can add substantial value to the high-throughput protocols.

**4.3 Reduction of time and cost**

An estimate of the time and cost savings using this combinatorial approach is presented in Table 4. The combinatorial approach is compared to the scenario where individual regular cubes would be manufactured and characterized for each column in the combinatorial sample. A feature of the combinatorial cube is that is takes up less material and space and can be built in a much shorter amount of time. The savings continues with sample preparation since sectioning and polishing can be achieved in units of twenty-four samples at a time. The savings for heat-treating or post-processing samples is based on using a small lab scale furnace that cannot process twenty-four regular samples at a time. For microscopy and hardness measurements, there is no savings for the actual measurement time; however, a significant amount of savings is gained by loading/setting up and unloading samples in units of twenty-four compared to one at a time. Additionally, the combinatorial sample allows for automating the measurement process in runs of twenty-four. Adding up the estimated time and cost savings in Table 4 gives a ratio of 3/16 (e.g., the combinatorial method would take three weeks compared to the regular sample taking 16 weeks). In other words, the combinatorial approach is approximately five times faster. This increase in productivity can be used to explore five times more process conditions or iterate five times faster. It is important to note that the information from the combinatorial sample is not equivalent to regular samples. The design and use of a combinatorial sample are driven by a need to gain insight quickly. Each design and application will be unique in the time and cost savings as well as the challenges in interpreting the results. These should be considered upfront to determine if this general approach should be used.

**Table 4 Estimated time and cost savings as the ratio of the combinatorial sample to the regular samples (i.e., cominbatorial samples result in a fraction of the effort compared to using regular samples).**

|  | Additive Material | Additive Process | Sample Preparation | Heat Treating | Microscopy | Hardness |
|---|---|---|---|---|---|---|
| Combi./Regular | 1/24 | 1/24 | 1/24 | 1/3 | 1/3 | 1/3 |

5. **Conclusions**

A high-throughput method for characterizing additively manufactured alloys was demonstrated through the manufacturer of combinatorial samples characterized by rapid measurement tools. This instance included five laser powers, four laser scan speeds, and three conditions (as-built, SR 800 °C for 1 hour, and SR 870 °C for 1 hour) built on a commercial L-PBF system using nickel superalloy 625. The conclusions are as follows:

1. Increasing laser scan speed decreases the grain size in the build plane for different laser powers with other parameters fixed (i.e., hatch spacing, power layer thickness, etc.). Increasing laser scan speed also produced a more random crystallographic texture along the build direction determined from EBSD. These observations using the combinatorial sample agree with the literature using regular samples, which show the utility of using the combinatorial sample for grain size and texture trends with laser processing parameters.

2. Stress-relief heat treatments had a more significant effect on hardness than laser power or scan speed. The hardness increased from the as-built condition for SR 800 °C due to precipitation hardening. The hardness subsequently decreased for SR 870 °C due to coarsening of precipitates along with a potential decrease in the ratio of $\gamma''$ to δ-phase. This conclusion is based on precipitate characterization from the literature for nickel superalloy 625 produced with similar AM parameters and post processes. The precipitation hardening during stress-relief is important for the mechanical properties of AM nickel superalloy 625.

3. Combinatorial AM samples that contain a library of different parameter combinations are a viable way to include a wider breadth of process and post-process parameters (five times more) without introducing significant time and cost. Repeats within the combinatorial sample are critical for interpreting trends and thermal models could help guide the design of combinatorial samples to estimate any neighbor effects. Additionally, the area or size of uniform conditions within the combinatorial sample is a potentially important variable evidenced from the difference in hardness between the combinatorial and regular sample.

**Acknowledgements**

The authors are indebted to a host across NIST who helped to figure out how to make good measurements despite the strange sample: Dr. Will Osborn and Ms. Maureen Williams with EBSD, Drs. Jarred Heigel and Jason Fox with LPBF, Dr. Kerry Siebein with XRD, Drs. Li-Piin Sung and Jae Hyun Kim with nanoindentation, and Dr. Mark Stoudt and Ms. Sandra Young for etching. We are also grateful for Dr. Fan Zhang for helping to understand the nanohardness trends, and discussions with Drs. Lyle Levine and Mark Stoudt on the process-structure relationships in AM nickel superalloy 625. Lastly, we appreciate discussions with Dr. Adam Creuziger about crystallographic texture measurements.

**Data Availability**

A significant amount of processed data is contained in the supplementary material. Additional raw/processed data required to reproduce these findings cannot be shared at this time due to technical or time limitations. Please contact the corresponding author for additional raw/processed data.

# Supplemental Material

**Title:** Demonstration of a laser powder bed fusion combinatorial sample for high-throughput microstructure and mechanical property characterization

**Authors:** Jordan S. Weaver[1,*], Adam L. Pintar[2], Carlos Beauchamp[3], Howie Joress[3], Kil-Won Moon[3], Thien Q. Phan[1]

[1]Engineering Laboratory, National Institute of Standards and Technology, 100 Bureau Drive, Gaithersburg, MD 20899; jordan.weaver@nist.gov, thien.phan@nist.gov
[2]Information Technology Laboratory, National Institute of Standards and Technology, 100 Bureau Drive, Gaithersburg, MD 20899, adam.pintar@nist.gov
[3]Materials Measurement Laboratory, National Institute of Standards and Technology, 100 Bureau Drive, Gaithersburg, MD 20899; carlos.beauchamp@nist.gov, howie.joress@nist.gov, kil-won.moon@nist.gov
[*]corresponding author

## S1. Additional AM settings process parameters

Table S1. Main process parameters. Default laser power and speed are 295 W and 960 mm s$^{-1}$, respectively. Uncertainties on the prescribed parameters are not available.

| Build Plate temperature (°C) | 80 | Hatch spacing (mm) | 0.11 |
|---|---|---|---|
| Recoating Blade | Ceramic | Stripe width (mm) | 10 |
| Powder layer thickness (mm) | 0.040 | Stripe rotation between layers | 67° rotation |
| Atmosphere | Argon | Nozzle Type | Standard |

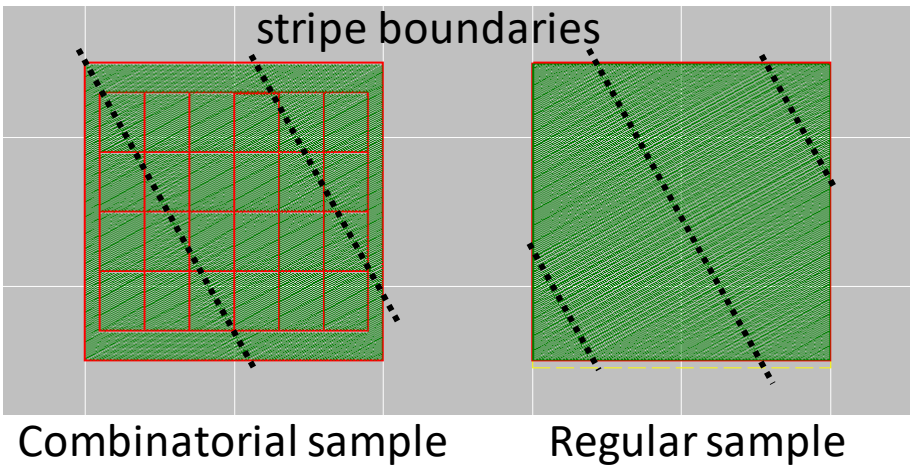

Fig. S1 Example of stripe boundaries for combinatorial and regular samples. Stripes are continuous across columns in the combinatorial sample because the stripe width (10 mm) and rotation strategy (67°) is the same for each column in the sample.

## S2. Chemical composition measurements

Table S2. Chemical composition measurements of a regular sample in the as-built condition. The composition meets ASTM F3056 - Standard Specification for Additive Manufacturing Nickel Alloy (UNS N06625) with Powder Bed Fusion [1]. Values are in weight percentage (wt. %). Elements were measured using inductively coupled plasma (ICP) except for N and O (measured using inert gas fusion) and C and S (measured using combustion analysis).

| Element | Standard Range (wt. %) | Measured (wt. %) |
| --- | --- | --- |
| Ni | Balance | Balance |
| Cr | 20.0 – 23.0 | 20.8 |
| Mo | 8.0 – 10.0 | 8.8 |
| Nb | 3.15 – 4.15 | 3.94 |
| Fe | 5.0 maximum | 0.8 |
| Ti | 0.4 maximum | 0.34 |
| Al | 0.4 maximum | 0.32 |
| Co |  | 0.19 |
| Si | 0.5 maximum | 0.08 |
| Mn | 0.5 maximum | 0.05 |
| C | 0.1 maximum | 0.03 |
| W |  | 0.03 |
| O |  | 0.019 |
| V |  | 0.014 |
| N |  | 0.009 |
| Cu |  | 0.006 |
| P | 0.015 maximum | 0.002 |
| S | 0.015 maximum | < 0.001 |

## S3. Nanoindentation area function calibrations on fused quartz

Table S3. Nanoindentation results on fused quartz averaged over a displacement range of 200 to 500 nm. Indenter modulus, indenter Poisson's ratio, and sample Poisson's ratio: $E_i = 1141\ GPa, v_i = 0.07, v_s = 0.18$, respectively

| Batch No. | Area Function ID | Avg. Modulus (GPa) | Std. Dev. (GPa) | Avg. Hardness (GPa) | Std. Dev. (GPa) | No. of Tests |
|---|---|---|---|---|---|---|
| 1 | A | 72.2 | 0.4 | 9.45 | 0.11 | 16 |
| 2 | A | 71.9 | 0.5 | 9.42 | 0.17 | 16 |
| 3 | B | 71.8 | 0.6 | 9.41 | 0.12 | 16 |
| 4 | C | 71.7 | 0.7 | 9.46 | 0.15 | 16 |
| 5 | C | 71.8 | 0.4 | 9.47 | 0.12 | 20 |
| 6 | C | 72.0 | 0.6 | 9.49 | 0.09 | 18 |

Table S4. Berkovich tip area function coefficients. The area, $A$, is a function of the contact depth, $h_c$: $A(h_c) = C_0 h_c^2 + C_1 h_c + C_2 h_c^{\frac{1}{2}} + C_3 h_c^{\frac{1}{4}}$. See Ref. [2]

| Area Function ID | $C_0$ | $C_1$ | $C_2$ | $C_3$ | Frame Stiffness (N/m) |
|---|---|---|---|---|---|
| A | 24.6657 | 142.519 | -796.916 | 87.6228 | 7.00264e+006 |
| B | 24.9149 | 299.102 | -148.436 | 256.099 | 7.47246e+006 |
| C | 24.4448 | 922.348 | -3382.58 | 2287.9 | 8.04089e+006 |

## S4. Nanoindentation results plotted versus position number

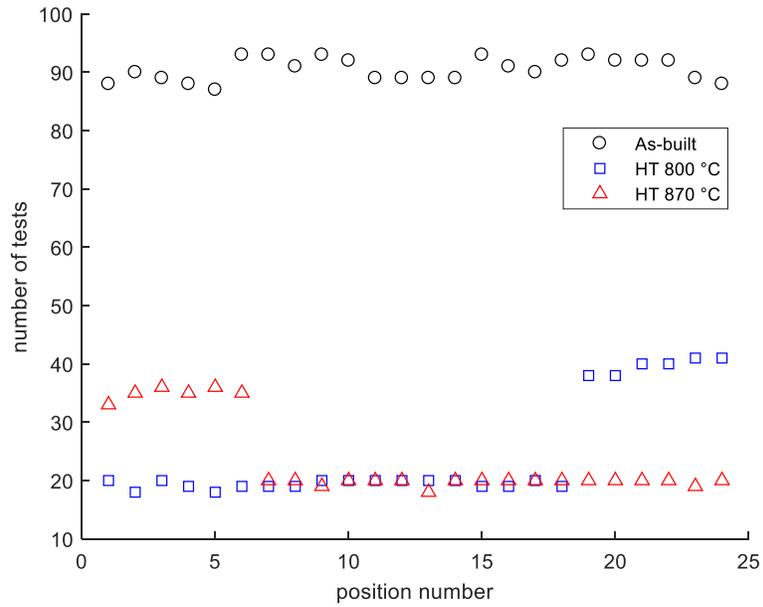

Fig. S2 Number of nanoindentation measurements per condition on combinatorial cubes. Position number refers to the 6 x 4 array of columns.

## S5. Optical micrograph and porosity area fraction measurements

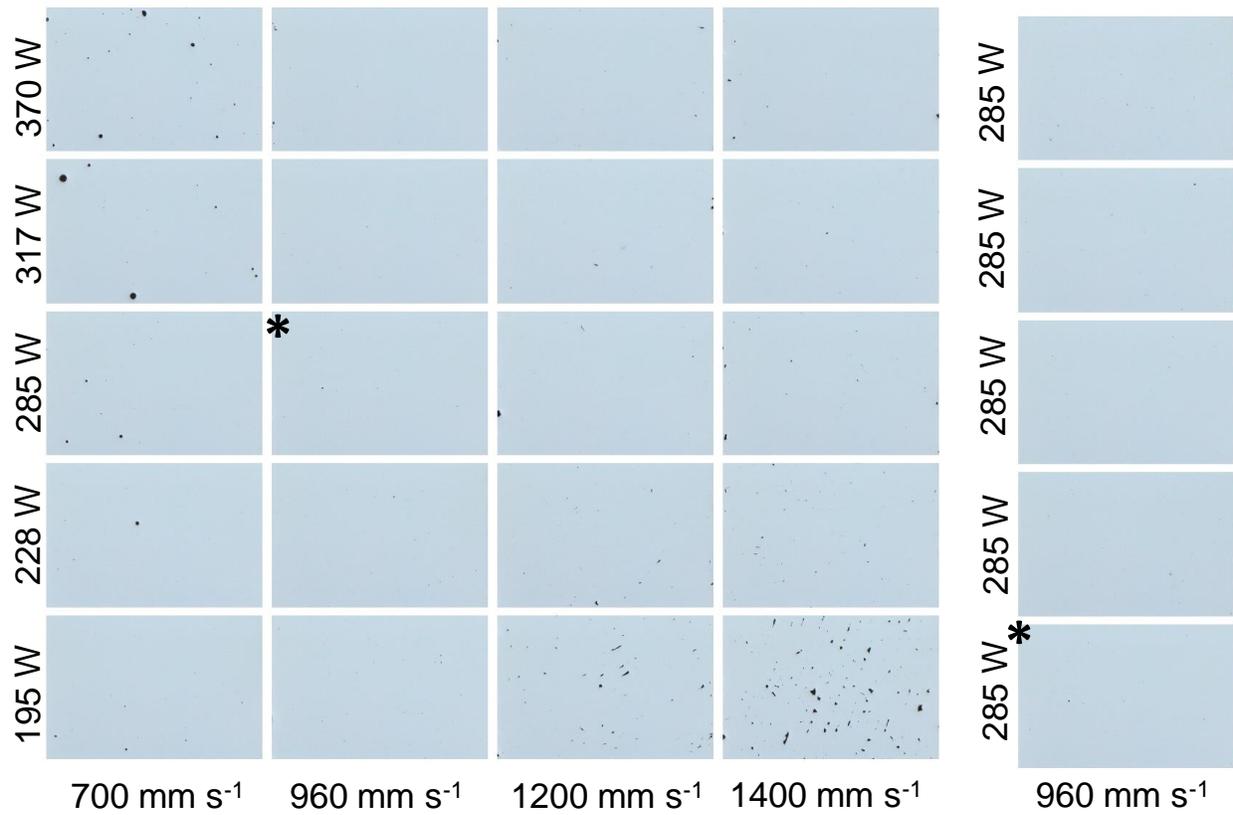

Fig. S3 Optical images of SR 800°C combinatorial sample. The field of view for each micrograph is 2.634 mm x 1.756 mm

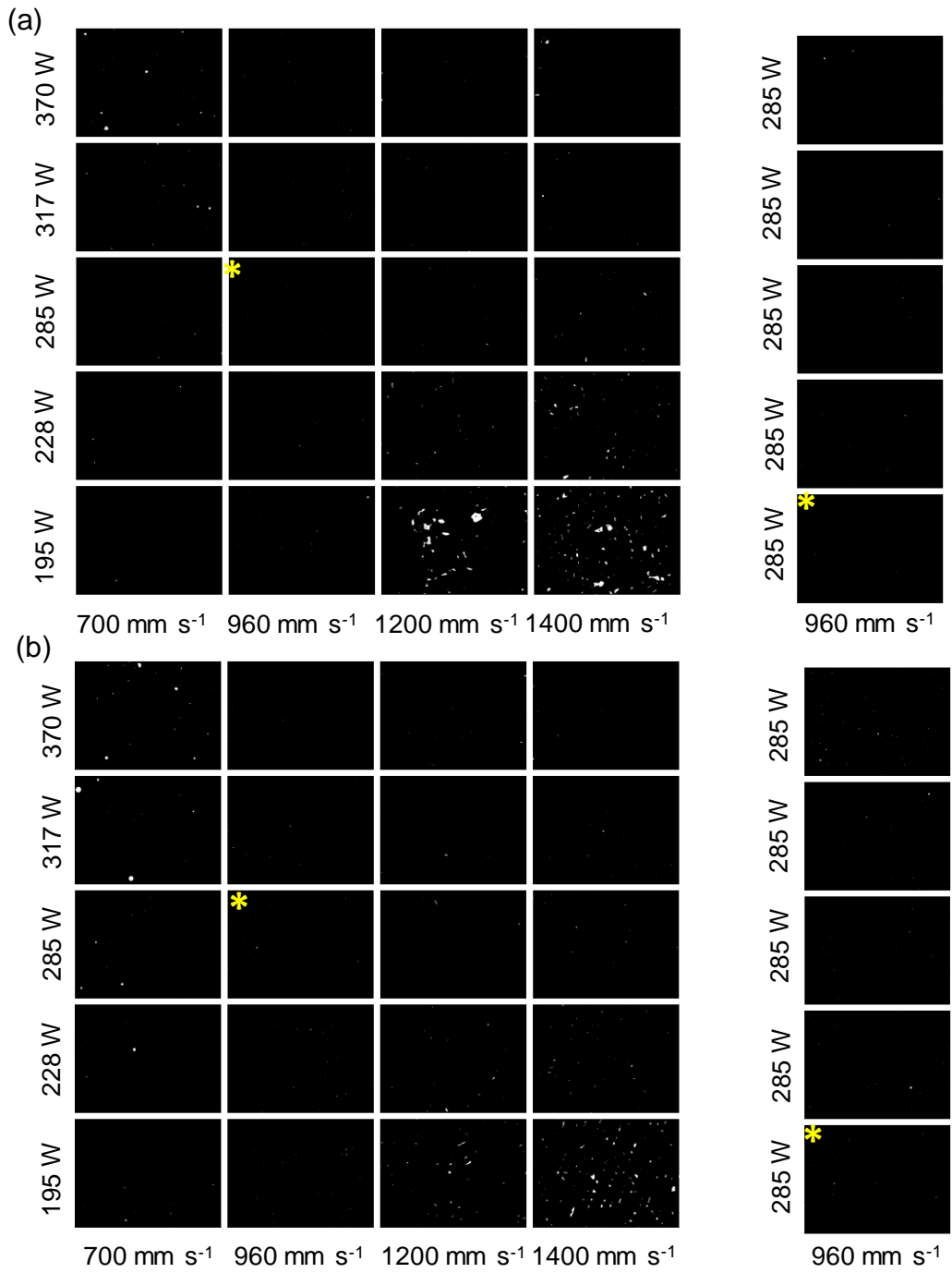

Fig. S4 Segmented optical micrographs with white regions for pores (a) As-built sample and (b) SR 800°C sample

Table S5 Porosity and pore circularity results per column

| Column No. | Laser Power (W) | Laser Scan Speed (mm/s) | Condition | Porosity (% Area) | Circularity |
|---|---|---|---|---|---|
| 1 | 285 | 960 | SR 800 °C | 0.07 | 0.95 |
| 2 | 195 | 1400 | SR 800 °C | 1.25 | 0.70 |
| 3 | 195 | 960 | SR 800 °C | 0.04 | 0.96 |
| 4 | 370 | 700 | SR 800 °C | 0.24 | 0.96 |
| 5 | 370 | 960 | SR 800 °C | 0.01 | 1.00 |
| 6 | 285 | 700 | SR 800 °C | 0.06 | 0.97 |
| 7 | 316 | 700 | SR 800 °C | 0.31 | 0.97 |
| 8 | 228 | 1400 | SR 800 °C | 0.11 | 0.85 |
| 9 | 285 | 1200 | SR 800 °C | 0.03 | 0.90 |
| 10 | 285 | 960 | SR 800 °C | 0.03 | 0.96 |
| 11 | 316 | 1400 | SR 800 °C | 0.03 | 0.91 |
| 12 | 316 | 1200 | SR 800 °C | 0.02 | 0.81 |
| 13 | 228 | 700 | SR 800 °C | 0.05 | 0.86 |
| 14 | 316 | 960 | SR 800 °C | 0.01 | 0.94 |
| 15 | 370 | 1200 | SR 800 °C | 0.02 | 0.90 |
| 16 | 285 | 960 | SR 800 °C | 0.02 | 0.94 |
| 17 | 285 | 1400 | SR 800 °C | 0.03 | 0.88 |
| 18 | 228 | 1200 | SR 800 °C | 0.09 | 0.82 |
| 19 | 285 | 960 | SR 800 °C | 0.04 | 0.94 |
| 20 | 370 | 1400 | SR 800 °C | 0.01 | 0.99 |
| 21 | 195 | 1200 | SR 800 °C | 0.30 | 0.74 |
| 22 | 195 | 700 | SR 800 °C | 0.03 | 0.93 |
| 23 | 228 | 960 | SR 800 °C | 0.03 | 0.96 |
| 24 | 285 | 960 | SR 800 °C | 0.02 | 0.99 |
| 1 | 285 | 960 | As-built | 0.02 | 0.94 |
| 2 | 195 | 1400 | As-built | 1.93 | 0.69 |
| 3 | 195 | 960 | As-built | 0.03 | 0.93 |
| 4 | 370 | 700 | As-built | 0.19 | 0.95 |
| 5 | 370 | 960 | As-built | 0.02 | 0.91 |
| 6 | 285 | 700 | As-built | 0.02 | 0.99 |
| 7 | 316 | 700 | As-built | 0.09 | 0.97 |
| 8 | 228 | 1400 | As-built | 0.42 | 0.77 |
| 9 | 285 | 1200 | As-built | 0.02 | 1.00 |
| 10 | 285 | 960 | As-built | 0.01 | 0.99 |
| 11 | 316 | 1400 | As-built | 0.03 | 0.94 |
| 12 | 316 | 1200 | As-built | 0.01 | 0.99 |
| 13 | 228 | 700 | As-built | 0.02 | 0.98 |
| 14 | 316 | 960 | As-built | 0.02 | 0.97 |
| 15 | 370 | 1200 | As-built | 0.01 | 0.94 |
| 16 | 285 | 960 | As-built | 0.01 | 1.00 |
| 17 | 285 | 1400 | As-built | 0.10 | 0.80 |
| 18 | 228 | 1200 | As-built | 0.12 | 0.71 |
| 19 | 285 | 960 | As-built | 0.02 | 0.98 |
| 20 | 370 | 1400 | As-built | 0.09 | 0.78 |
| 21 | 195 | 1200 | As-built | 1.75 | 0.66 |
| 22 | 195 | 700 | As-built | 0.01 | 1.00 |
| 23 | 228 | 960 | As-built | 0.01 | 1.00 |
| 24 | 285 | 960 | As-built | 0.00 | 0.95 |
| Regular | 285 | 960 | As-built | 0.01 | 1.00 |
| Regular | 285 | 960 | As-built | 0.02 | 0.96 |
| Regular | 285 | 960 | As-built | 0.01 | 0.99 |
| Regular | 285 | 960 | As-built | 0.01 | 0.98 |
| Regular | 285 | 960 | As-built | 0.01 | 0.99 |

## S6. EBSD and grain size measurements

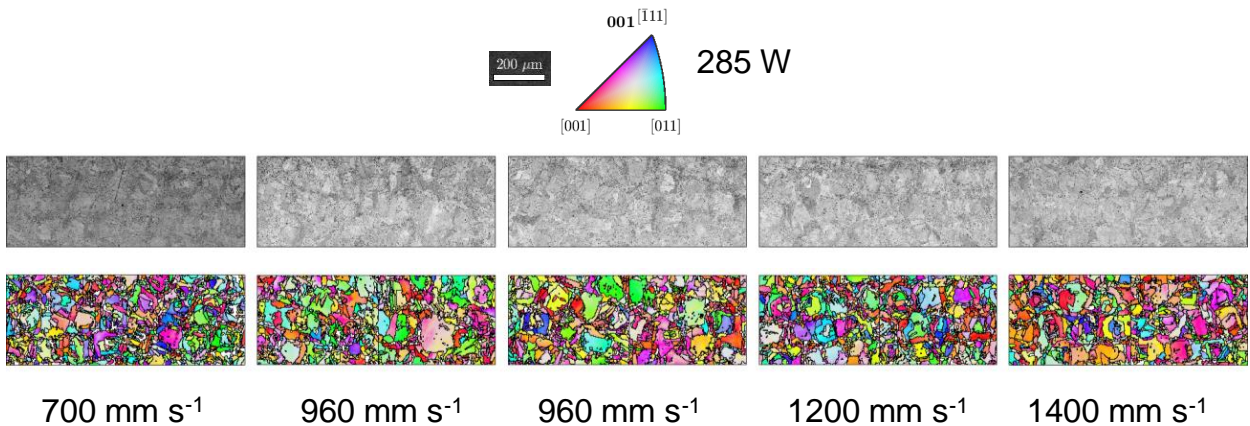

Fig. S5 EBSD band contrast and inverse pole figure maps for several 285 W conditions on the SR 870°C sample.

Table S6a Grain size measurements from EBSD. The 5[th] % is unreliable because it is very close to the minimum detectable grain size. Grain sizes are the equivalent circular area radius in units of micrometers.

| Condition | Power (W) | Speed (mm s$^{-1}$) | P/S ratio (W mm s$^{-1}$) | 5[th] % | 50[th] % | 95[th] % | Mean | No. Grains | % not-indexed | % cleanup | Texture Index | Entropy |
|---|---|---|---|---|---|---|---|---|---|---|---|---|
| As-built | 285 | 960 | 0.297 | 5.1 | 8.0 | 25.0 | 10.6 | 613 | 2 | 2 | 1.7 | -0.32 |
| | 195 | 1400 | 0.139 | 4.9 | 6.5 | 13.3 | 7.4 | 1020 | 9 | 4 | 1.1 | -0.05 |
| | 195 | 960 | 0.203 | 5.0 | 7.6 | 19.8 | 9.3 | 830 | 2 | 2 | 1.1 | -0.06 |
| | 370 | 700 | 0.529 | 5.0 | 8.0 | 25.0 | 10.2 | 669 | 2 | 2 | 1.3 | -0.15 |
| | 370 | 960 | 0.385 | 5.1 | 8.3 | 24.4 | 10.4 | 649 | 2 | 2 | 1.6 | -0.28 |
| | 285 | 700 | 0.407 | 5.1 | 8.4 | 24.5 | 10.8 | 601 | 2 | 2 | 1.4 | -0.20 |
| | 316 | 700 | 0.451 | 5.1 | 8.0 | 25.0 | 10.3 | 650 | 2 | 1 | 1.5 | -0.25 |
| | 228 | 1400 | 0.163 | 4.9 | 6.8 | 15.8 | 8.0 | 1025 | 3 | 2 | 1.1 | -0.03 |
| | 285 | 1200 | 0.238 | 5.1 | 7.9 | 22.8 | 9.9 | 721 | 2 | 2 | 1.2 | -0.10 |
| | 285 | 960 | 0.297 | 5.1 | 8.2 | 23.2 | 10.4 | 658 | 2 | 1 | 1.6 | -0.29 |
| | 316 | 1400 | 0.226 | 4.9 | 7.3 | 17.2 | 8.7 | 957 | 1 | 1 | 1.1 | -0.04 |
| | 316 | 1200 | 0.263 | 4.9 | 7.7 | 22.7 | 9.8 | 731 | 1 | 1 | 1.4 | -0.17 |
| | 228 | 700 | 0.326 | 5.1 | 8.1 | 26.0 | 10.7 | 609 | 1 | 1 | 1.6 | -0.25 |
| | 316 | 960 | 0.329 | 5.1 | 7.9 | 25.4 | 10.4 | 631 | 1 | 1 | 2.3 | -0.50 |
| | 370 | 1200 | 0.308 | 4.9 | 7.6 | 22.1 | 9.5 | 783 | 1 | 1 | 1.3 | -0.15 |
| | 285 | 960 | 0.297 | 4.9 | 8.4 | 23.7 | 10.7 | 621 | 1 | 1 | 1.5 | -0.24 |
| | 285 | 1400 | 0.204 | 5.0 | 7.3 | 18.0 | 8.7 | 924 | 2 | 2 | 1.1 | -0.05 |
| | 228 | 1200 | 0.190 | 5.0 | 7.1 | 15.5 | 8.3 | 1016 | 3 | 2 | 1.1 | -0.03 |
| | 285 | 960 | 0.297 | 5.1 | 8.0 | 23.4 | 10.4 | 657 | 2 | 2 | 1.4 | -0.20 |
| | 370 | 1400 | 0.264 | 4.9 | 7.5 | 18.4 | 9.0 | 895 | 2 | 2 | 1.2 | -0.09 |
| | 195 | 1200 | 0.163 | 5.0 | 7.1 | 14.8 | 8.1 | 1052 | 4 | 3 | 1.1 | -0.03 |
| | 195 | 700 | 0.279 | 5.1 | 8.4 | 28.4 | 11.1 | 569 | 1 | 1 | 1.4 | -0.20 |
| | 228 | 960 | 0.238 | 5.1 | 8.4 | 21.9 | 10.4 | 671 | 2 | 2 | 1.4 | -0.17 |
| | 285 | 960 | 0.297 | 4.8 | 8.3 | 25.2 | 10.5 | 638 | 1 | 1 | 1.5 | -0.24 |
| SR 870°C | 285 | 960 | 0.297 | 4.9 | 7.9 | 22.8 | 10.2 | 650 | <0.5 | <0.5 | 1.3 | -0.17 |
| | 285 | 700 | 0.407 | 4.9 | 8.2 | 22.4 | 10.3 | 674 | <0.5 | <0.5 | 1.5 | -0.18 |
| | 285 | 1200 | 0.238 | 4.8 | 7.9 | 21.2 | 9.6 | 762 | <0.5 | <0.5 | 1.2 | -0.11 |
| | 285 | 960 | 0.297 | 4.9 | 7.9 | 23.8 | 10.2 | 649 | <0.5 | <0.5 | 1.3 | -0.17 |
| | 285 | 1400 | 0.204 | 5.1 | 7.3 | 18.4 | 9.0 | 864 | <0.5 | <0.5 | 1.1 | -0.06 |

Table S6b Grain size analysis using 5° instead of 3° misorientation angle to define grains. The 5th % is unreliable because it is very close to the minimum detectable grain size. Grain sizes are the equivalent circular area radius in units of micrometers.

| Condition | Power (W) | Speed (mm s$^{-1}$) | P/S ratio (W mm s$^{-1}$) | 5th % | 50th % | 95th % | Mean | No. Grains |
|---|---|---|---|---|---|---|---|---|
| As-built | 285 | 960 | 0.297 | 5.1 | 8.4 | 26.3 | 11.4 | 529 |
| | 195 | 1400 | 0.139 | 4.9 | 6.5 | 13.4 | 7.5 | 1032 |
| | 195 | 960 | 0.203 | 5.1 | 8.0 | 21.0 | 9.7 | 786 |
| | 370 | 700 | 0.529 | 5.1 | 8.4 | 26.4 | 10.7 | 611 |
| | 370 | 960 | 0.385 | 5.1 | 8.8 | 27.4 | 11.0 | 576 |
| | 285 | 700 | 0.407 | 5.1 | 8.7 | 25.5 | 11.2 | 560 |
| | 316 | 700 | 0.451 | 5.1 | 8.3 | 26.0 | 10.7 | 609 |
| | 228 | 1400 | 0.163 | 4.9 | 6.8 | 16.4 | 8.2 | 1014 |
| | 285 | 1200 | 0.238 | 5.1 | 7.9 | 23.4 | 10.1 | 701 |
| | 285 | 960 | 0.297 | 5.1 | 8.8 | 26.3 | 11.1 | 574 |
| | 316 | 1400 | 0.226 | 4.9 | 7.5 | 17.6 | 8.9 | 926 |
| | 316 | 1200 | 0.263 | 4.9 | 7.8 | 24.7 | 10.1 | 680 |
| | 228 | 700 | 0.326 | 5.1 | 8.6 | 28.1 | 11.4 | 535 |
| | 316 | 960 | 0.329 | 5.1 | 8.5 | 28.8 | 11.3 | 538 |
| | 370 | 1200 | 0.308 | 5.0 | 7.7 | 22.7 | 9.8 | 743 |
| | 285 | 960 | 0.297 | 5.1 | 9.1 | 26.4 | 11.5 | 538 |
| | 285 | 1400 | 0.204 | 5.0 | 7.4 | 18.4 | 9.0 | 899 |
| | 228 | 1200 | 0.190 | 5.1 | 7.3 | 15.7 | 8.5 | 986 |
| | 285 | 960 | 0.297 | 5.1 | 8.4 | 25.3 | 11.0 | 589 |
| | 370 | 1400 | 0.264 | 4.9 | 7.6 | 19.2 | 9.2 | 863 |
| | 195 | 1200 | 0.163 | 5.1 | 7.2 | 14.9 | 8.2 | 1041 |
| | 195 | 700 | 0.279 | 5.0 | 8.8 | 29.1 | 11.6 | 515 |
| | 228 | 960 | 0.238 | 5.1 | 8.8 | 24.8 | 11.2 | 592 |
| | 285 | 960 | 0.297 | 4.9 | 8.5 | 26.7 | 11.0 | 585 |

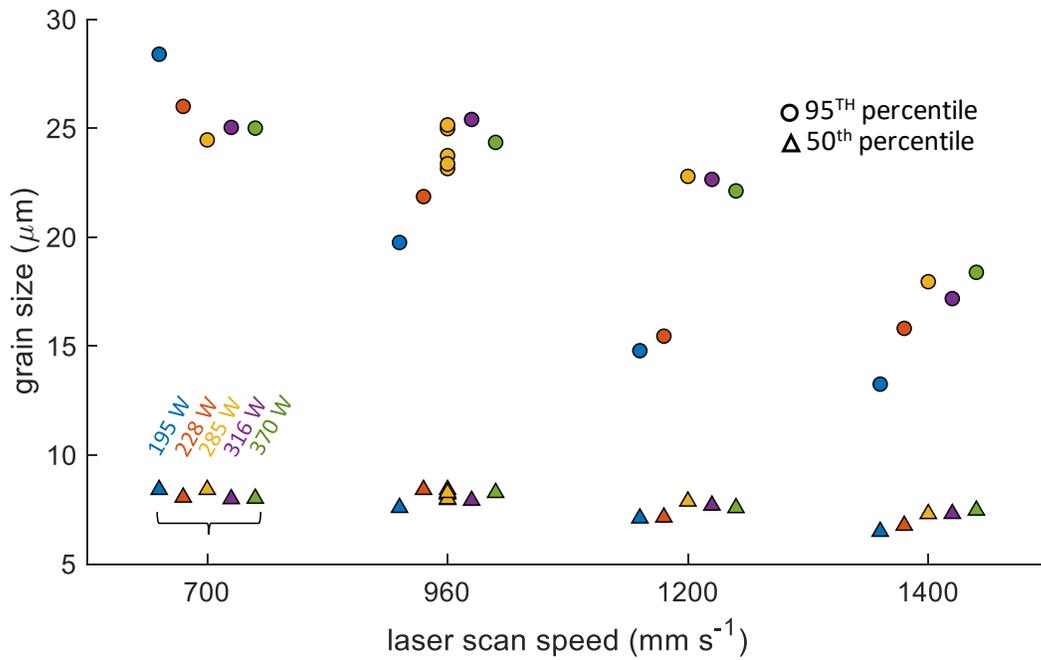

Fig. S6 Grain size measurements for the mean and 95$^{th}$ percentile versus laser scan speed color coded based on laser power.

**S8. Indentation measurements**

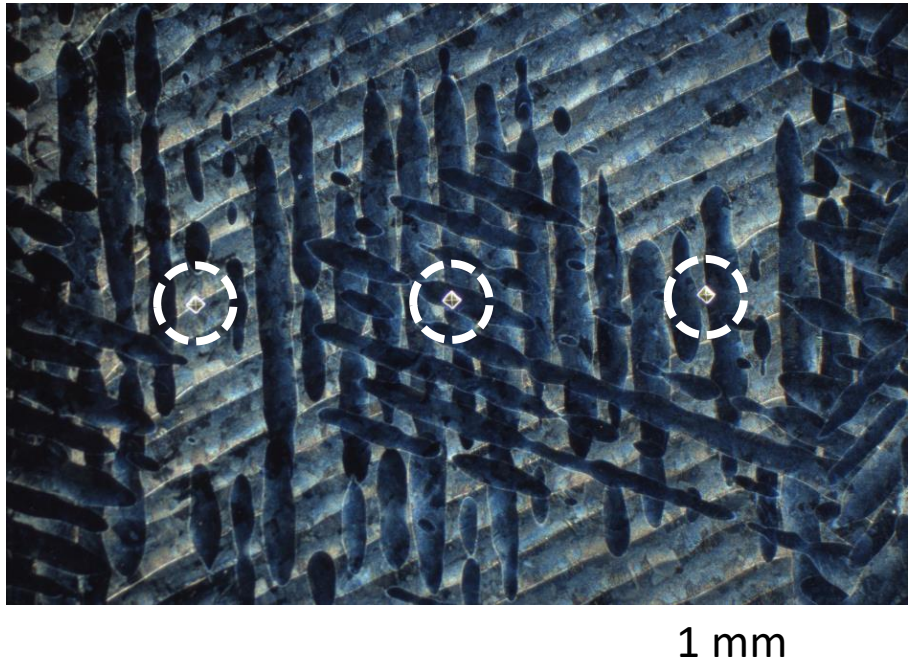

Fig. S7 Dark field optical micrograph of position 1 for the as-built combinatorial sample showing Vickers indents. The sample was etched with aqua regia to reveal the melt pool boundaries.

Table S7. Indentation measurements for 60 combinations of laser power, scan speed and condition. The columns for each measurement are the average, lower bound, and upper bound from left to right. The bounds are for $p = 95$ %.

| Laser Power (W) | Scan Speed (mm s$^{-1}$) | Condition | Modulus (GPa) | | | Nanohardness (GPa) | | | Vickers Hardness (HV0.5) | | |
|---|---|---|---|---|---|---|---|---|---|---|---|
| 195 | 700 | SR 800 °C | 218.0 | 209.4 | 226.9 | 4.79 | 4.66 | 4.94 | 315.0 | 300.5 | 327.7 |
| 195 | 960 | SR 800 °C | 219.5 | 208.6 | 230.0 | 4.74 | 4.57 | 4.91 | 320.7 | 307.6 | 334.8 |
| 195 | 1200 | SR 800 °C | 220.6 | 211.2 | 229.0 | 4.79 | 4.64 | 4.93 | 312.3 | 298.8 | 325.3 |
| 195 | 1400 | SR 800 °C | 221.0 | 211.5 | 231.1 | 4.79 | 4.61 | 4.95 | 311.3 | 298.4 | 324.2 |
| 228 | 700 | SR 800 °C | 224.3 | 213.7 | 234.5 | 4.85 | 4.67 | 5.03 | 318.0 | 304.2 | 330.7 |
| 228 | 960 | SR 800 °C | 216.1 | 206.5 | 225.7 | 4.75 | 4.59 | 4.91 | 322.5 | 309.9 | 336.9 |
| 228 | 1200 | SR 800 °C | 216.5 | 206.5 | 226.9 | 4.76 | 4.59 | 4.93 | 317.7 | 302.4 | 330.7 |
| 228 | 1400 | SR 800 °C | 218.2 | 208.6 | 228.1 | 4.76 | 4.59 | 4.92 | 315.8 | 301.9 | 329.6 |
| 285 | 700 | SR 800 °C | 212.4 | 202.6 | 222.9 | 4.80 | 4.64 | 4.97 | 318.9 | 304.8 | 332.6 |
| 285 | 960 | SR 800 °C | 217.8 | 212.0 | 223.8 | 4.81 | 4.73 | 4.88 | 321.8 | 315.9 | 327.8 |
| 285 | 1200 | SR 800 °C | 217.8 | 208.3 | 227.7 | 4.79 | 4.62 | 4.96 | 321.3 | 307.9 | 335.0 |
| 285 | 1400 | SR 800 °C | 218.0 | 208.9 | 228.3 | 4.78 | 4.62 | 4.95 | 318.8 | 305.0 | 331.6 |
| 316 | 700 | SR 800 °C | 210.9 | 201.2 | 221.4 | 4.82 | 4.65 | 4.98 | 315.4 | 302.1 | 328.8 |
| 316 | 960 | SR 800 °C | 223.3 | 213.4 | 233.2 | 4.87 | 4.70 | 5.04 | 328.2 | 315.3 | 341.7 |
| 316 | 1200 | SR 800 °C | 220.2 | 210.3 | 229.9 | 4.82 | 4.65 | 4.99 | 325.3 | 312.1 | 340.2 |
| 316 | 1400 | SR 800 °C | 222.4 | 212.0 | 232.6 | 4.88 | 4.72 | 5.05 | 325.0 | 312.0 | 338.0 |
| 370 | 700 | SR 800 °C | 216.1 | 205.5 | 226.8 | 4.83 | 4.66 | 5.01 | 323.0 | 309.3 | 336.5 |
| 370 | 960 | SR 800 °C | 213.6 | 202.8 | 223.5 | 4.89 | 4.73 | 5.06 | 323.4 | 310.3 | 337.8 |
| 370 | 1200 | SR 800 °C | 222.5 | 213.0 | 232.2 | 4.84 | 4.68 | 5.01 | 332.7 | 319.5 | 346.8 |
| 370 | 1400 | SR 800 °C | 221.1 | 212.4 | 231.5 | 4.80 | 4.64 | 4.97 | 320.4 | 306.7 | 334.8 |
| 195 | 700 | SR 870 °C | 209.6 | 199.3 | 219.6 | 4.47 | 4.31 | 4.64 | 290.3 | 277.1 | 304.1 |
| 195 | 960 | SR 870 °C | 217.3 | 208.6 | 227.1 | 4.55 | 4.39 | 4.71 | 298.8 | 285.8 | 313.2 |
| 195 | 1200 | SR 870 °C | 218.9 | 209.1 | 229.6 | 4.59 | 4.43 | 4.77 | 299.0 | 285.3 | 312.2 |
| 195 | 1400 | SR 870 °C | 214.8 | 205.1 | 224.3 | 4.57 | 4.41 | 4.73 | 275.7 | 262.3 | 289.0 |
| 228 | 700 | SR 870 °C | 215.5 | 205.1 | 225.4 | 4.55 | 4.38 | 4.72 | 299.0 | 285.8 | 312.2 |
| 228 | 960 | SR 870 °C | 212.4 | 202.7 | 222.9 | 4.48 | 4.30 | 4.65 | 300.7 | 287.7 | 314.0 |
| 228 | 1200 | SR 870 °C | 223.0 | 212.9 | 232.6 | 4.56 | 4.40 | 4.73 | 298.0 | 283.7 | 312.5 |
| 228 | 1400 | SR 870 °C | 221.6 | 212.3 | 231.4 | 4.67 | 4.50 | 4.83 | 288.3 | 275.5 | 300.7 |
| 285 | 700 | SR 870 °C | 217.3 | 208.1 | 227.6 | 4.51 | 4.36 | 4.68 | 297.1 | 283.8 | 311.5 |
| 285 | 960 | SR 870 °C | 215.5 | 209.9 | 220.9 | 4.53 | 4.46 | 4.61 | 294.9 | 289.3 | 300.9 |
| 285 | 1200 | SR 870 °C | 219.7 | 209.6 | 229.9 | 4.59 | 4.41 | 4.77 | 297.2 | 283.4 | 311.1 |
| 285 | 1400 | SR 870 °C | 221.2 | 211.1 | 231.9 | 4.67 | 4.50 | 4.83 | 301.5 | 288.0 | 314.1 |
| 316 | 700 | SR 870 °C | 214.9 | 205.2 | 225.1 | 4.52 | 4.34 | 4.69 | 295.2 | 281.3 | 308.7 |
| 316 | 960 | SR 870 °C | 216.4 | 206.3 | 226.0 | 4.63 | 4.45 | 4.79 | 296.5 | 283.3 | 310.3 |
| 316 | 1200 | SR 870 °C | 213.5 | 204.1 | 223.8 | 4.62 | 4.45 | 4.79 | 300.8 | 286.7 | 313.8 |
| 316 | 1400 | SR 870 °C | 218.7 | 209.3 | 228.9 | 4.63 | 4.46 | 4.81 | 303.6 | 290.4 | 317.0 |
| 370 | 700 | SR 870 °C | 215.0 | 204.9 | 225.3 | 4.52 | 4.36 | 4.68 | 292.2 | 278.3 | 305.7 |
| 370 | 960 | SR 870 °C | 219.0 | 208.8 | 228.5 | 4.58 | 4.42 | 4.74 | 295.6 | 282.2 | 309.0 |
| 370 | 1200 | SR 870 °C | 221.0 | 211.4 | 230.6 | 4.60 | 4.44 | 4.76 | 298.5 | 285.2 | 312.4 |
| 370 | 1400 | SR 870 °C | 217.3 | 207.3 | 226.8 | 4.57 | 4.40 | 4.75 | 300.8 | 288.0 | 314.4 |
| 195 | 700 | As-built | 204.4 | 196.4 | 212.0 | 4.09 | 3.98 | 4.21 | 286.8 | 272.9 | 300.0 |
| 195 | 960 | As-built | 207.7 | 199.9 | 215.0 | 4.20 | 4.09 | 4.33 | 293.5 | 280.1 | 307.0 |
| 195 | 1200 | As-built | 208.0 | 199.7 | 216.2 | 4.11 | 4.00 | 4.24 | 284.1 | 269.3 | 298.7 |
| 195 | 1400 | As-built | 206.5 | 198.6 | 214.5 | 4.23 | 4.11 | 4.34 | 278.4 | 265.1 | 292.4 |
| 228 | 700 | As-built | 210.6 | 202.7 | 218.7 | 4.23 | 4.12 | 4.35 | 295.8 | 282.7 | 309.2 |
| 228 | 960 | As-built | 205.0 | 196.5 | 212.9 | 4.12 | 4.01 | 4.24 | 284.3 | 270.5 | 298.1 |
| 228 | 1200 | As-built | 211.9 | 203.9 | 219.4 | 4.22 | 4.11 | 4.34 | 294.7 | 281.6 | 306.5 |
| 228 | 1400 | As-built | 208.7 | 200.7 | 216.2 | 4.13 | 4.01 | 4.25 | 293.3 | 280.7 | 306.0 |
| 285 | 700 | As-built | 208.1 | 200.1 | 216.9 | 4.29 | 4.18 | 4.41 | 288.4 | 274.2 | 302.5 |
| 285 | 960 | As-built | 205.7 | 201.1 | 209.8 | 4.19 | 4.14 | 4.24 | 291.4 | 285.5 | 297.1 |
| 285 | 1200 | As-built | 207.7 | 199.7 | 215.5 | 4.16 | 4.05 | 4.26 | 301.2 | 287.6 | 314.6 |
| 285 | 1400 | As-built | 211.7 | 203.6 | 219.6 | 4.28 | 4.17 | 4.40 | 291.6 | 278.3 | 304.3 |
| 316 | 700 | As-built | 207.4 | 198.5 | 215.9 | 4.26 | 4.14 | 4.37 | 295.3 | 281.8 | 308.4 |
| 316 | 960 | As-built | 208.1 | 200.2 | 216.1 | 4.28 | 4.16 | 4.39 | 302.1 | 288.9 | 315.0 |
| 316 | 1200 | As-built | 208.2 | 200.7 | 216.2 | 4.27 | 4.16 | 4.38 | 292.3 | 278.9 | 306.3 |
| 316 | 1400 | As-built | 211.1 | 203.4 | 219.2 | 4.29 | 4.17 | 4.40 | 291.3 | 277.2 | 304.3 |
| 370 | 700 | As-built | 205.2 | 196.9 | 213.7 | 4.08 | 3.97 | 4.19 | 288.5 | 274.5 | 301.0 |
| 370 | 960 | As-built | 206.6 | 198.1 | 214.5 | 4.22 | 4.10 | 4.33 | 293.6 | 281.2 | 306.6 |
| 370 | 1200 | As-built | 209.3 | 201.5 | 217.2 | 4.23 | 4.12 | 4.34 | 298.5 | 286.3 | 311.9 |
| 370 | 1400 | As-built | 210.2 | 202.6 | 218.4 | 4.25 | 4.13 | 4.36 | 282.5 | 269.5 | 295.6 |

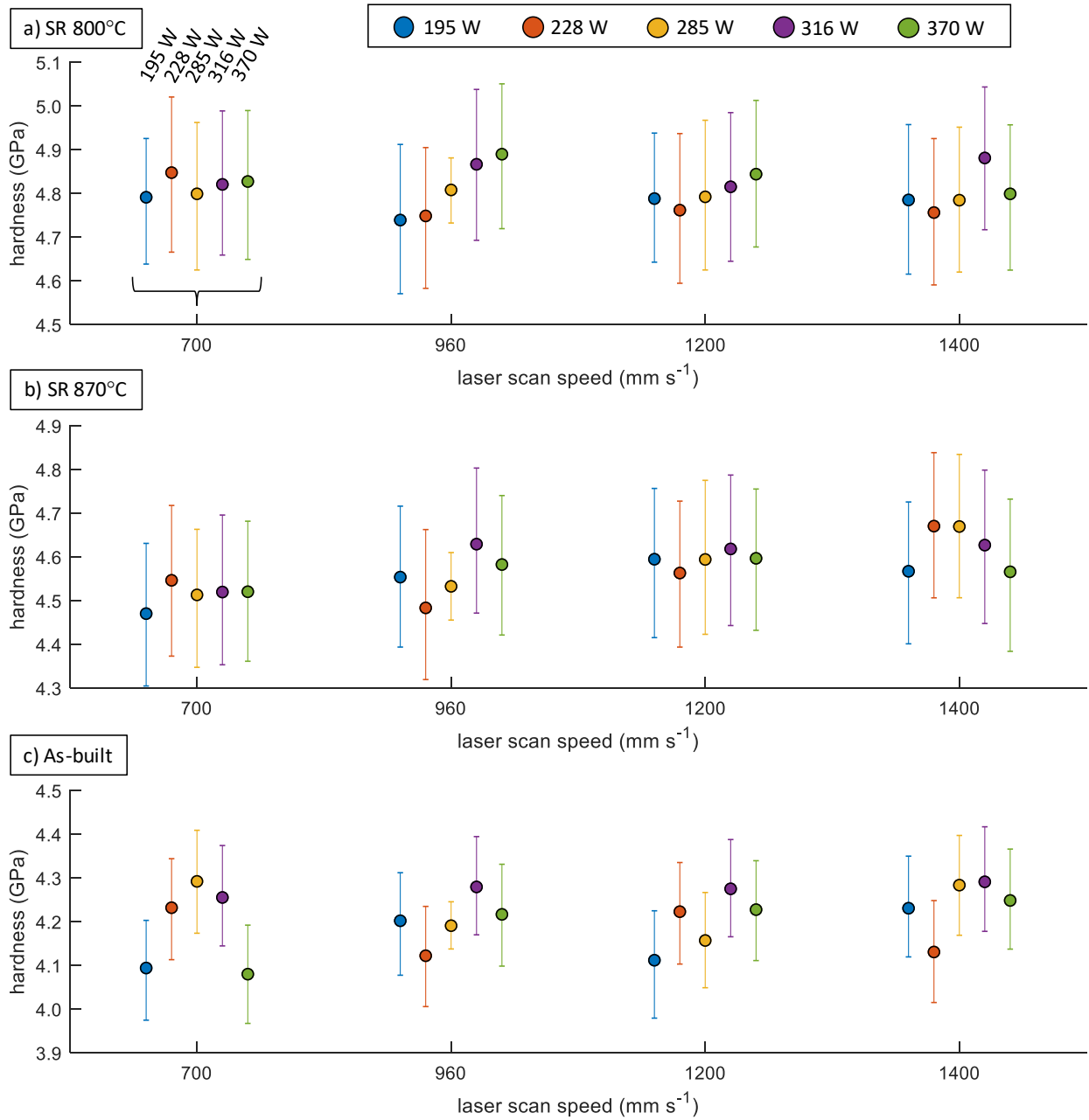

Fig. S8 Nanoindentation hardness grouped by laser scan speed for all three conditions and color coded according to laser power.

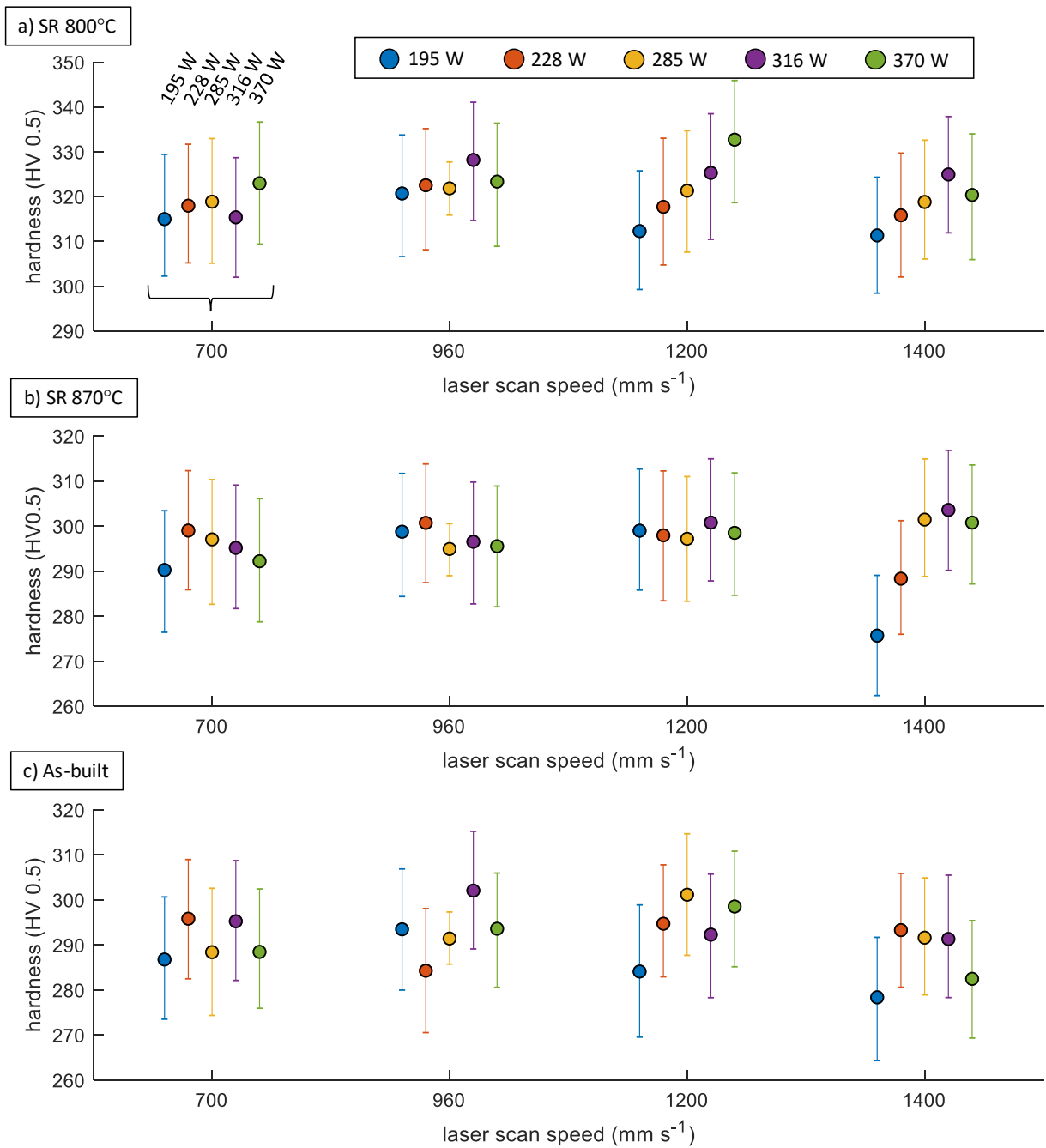

Fig. S9. Microindentation Vickers hardness grouped by laser scan speed for all three conditions and color coded according to laser power.